\documentclass{aa}  
\usepackage{graphicx}
\usepackage{graphicx}
\usepackage{verbatim}
\usepackage{subcaption}
\usepackage{txfonts}
\usepackage{xcolor}
\newcommand{\rhill}{\ensuremath{R_{\textrm{Hill}}}}

\begin{document} 

    \title{Machine learning approach for mapping the stable orbits around planets}
    \titlerunning{ML for mapping stability regions}


   \author{Tiago F. L. L. Pinheiro
          \inst{1},
          Rafael Sfair\inst{1,2, 3}
          \and
          Giovana Ramon\inst{1}
          }

    \institute{Grupo de Din\^amica Orbital e Planetologia, S\~ao Paulo State University, UNESP,
               Guaratinguet\'a, CEP 12516-410, S\~ao Paulo, Brazil\\
              \email{francisco.pinheiro@unesp.br}
          \and
             Eberhard Karls Universität Tübingen,
             Auf der Morgenstelle, 10, 72076, Tübingen, Germany
            \and
            LESIA, Observatoire de Paris, Université PSL, CNRS, Sorbonne Université, 5 place Jules Janssen, 92190 Meudon, France
             }

      \date{Received \today; accepted ---}

   \abstract
 {Numerical N-body simulations are typically employed 
 to map stability regions around exoplanets,
providing insights into the potential presence 
of satellites and ring systems.}
   {Use Machine Learning (ML) techniques to generate predictive maps of stable
    regions surrounding hypothetical planet, and
    this approach can also be applied to planet-satellite systems, 
    planetary ring systems, and other similar systems.}
   {From a set of $10^5$ numerical simulations, each 
   incorporating nine orbital features for the planet and 
   test particle, we created a comprehensive dataset of three-body
   problem outcomes (star-planet-test particle). 
   Simulations were classified as stable or unstable, 
   considering the stability criteria that a 
   particle must remain stable over a time span of  10$^{4}$ 
   orbital periods of the planet.
   Various ML algorithms were compared and fine-tuned through
   hyperparameter optimization to identify the most effective predictive 
   model, and all the tree based algorithms demonstrated comparable accuracy 
   performance.}
   {The optimal model, employing the Extreme Gradient Boosting (XGBoost) algorithm, achieved 
   an accuracy of 98.48\% ,
    with 94\% recall and precision for stable particles and 99\% 
    for unstable particles.}
   {ML algorithms significantly reduce computational 
   time in three-body simulations, achieving speeds approximately $10^{5}$ 
   times faster than traditional numerical simulations. 
   From saved training models, predictions of entire stability maps are made in less than a second, 
   while an equivalent numerical simulation can take up to a few days.
   Our ML model results will be accessible
   through a forthcoming public web interface, 
   facilitating broader scientific application.}


   \maketitle


\section{Introduction}
\label{Introduction}

The population of known exoplanets has grown year-over-year, 
prompting extensive research into the stability of extrasolar planetary systems. 
To investigate the potential existence of satellites or ring systems around 
these exoplanets, numerical N-body simulations are frequently employed. 
These simulations map stability regions around a planet by utilizing 
a grid of orbital parameters as initial conditions
for the N-body integration.

Several studies have expanded this analysis to various contexts in
different gravitational settings.
\citet{Hunter1967} examined the stability of satellites orbiting Jupiter under
the influence of solar gravity. Their findings indicate that the stable zone
for prograde motion extends to 0.45 of the Hill radius ($r_\mathrm{Hill}$), while for
retrograde motion, it extends to 0.75 $r_\mathrm{Hill}$. 
\citet{Vieira2001} investigated the gravitational capture in the Sun-Uranus-satellite system.
They employed a diagram of initial conditions of semi-major axis versus eccentricity
to identify stable regions around Uranus, considering particles with various
initial inclinations and arguments of pericenter.

\citet{Holman1999} numerically investigated the stability of S-type 
and P-type planet orbit around a 
binary star system
where the planet is treated as a test particle.
Their work resulted in an empirical expression
for the critical semi-major axis of a stable particle initially in a circular
and planar orbit, relating it to binary distance, mass ratio, and eccentricity.
The authors analyzed a range of binary mass ratios $\mu$ = [0-0.9] and
eccentricities e~~=~[0-0.8].

\citet{Domingos2006} examined the stability of hypothetical satellites
orbiting extrasolar planets with a mass ratio $\mu = 10^{-3}$. They derived an
empirical expression for the stability boundary in both prograde ($i=0^\circ$) and
retrograde ($i=180^\circ$) orbits, formulated in terms of the Hill radius and the
eccentricities of the planet and the satellite.

\citet{Rieder2016} conducted N-body simulations to analyze the stability of a
potential ring system around the candidate planet J1407b, thereby imposing constraints
on both the orbital and physical parameters of this system. For the PDS110b system, 
\citet{Pinheiro2021} performed millions of numerical simulations of the three-body problem, 
aiming to refine parameters that remain undefined based on observational data, including the 
mass and eccentricity of unseen secondary companion, 
as well as the ring inclination and radial size.

While the aforementioned studies have provided valuable insights, conducting
such numerical simulations for many systems incurs substantial computational
expense. To address this challenge, a novel method to expedite this process involves the
use of ML algorithms. 
In this approach, N-body simulations generate a dataset
for training ML models, which can 
then predict outcomes for unseen data more efficiently.

The identification of stability regions using ML methods has
already been demonstrated in previous studies. \citet{Tamayo2016} applied
ML algorithms to predict the stability of tightly packed planetary
systems, consisting of three planets 
orbiting a central star. Their models achieved 
a predictive accuracy of $90\%$ over the test set, 
and were three orders of 
magnitude faster than direct N-body simulations.

\citet{Tamayo2020} developed SPOCK 
(Stability of Planetary Orbital Configurations Classifier), 
a ML model capable of predicting the stability 
of compact systems with three or more planets. 
This approach is 10$^{5}$ times faster than numerical simulations.

\citet{Cranmer2021} introduced a probabilistic ML model that predicts the stability of 
compact multi-planet systems with three or more planets, 
including when these systems are likely to become unstable. 
This model demonstrates over two orders of magnitude greater 
accuracy in predicting instability times compared to analytical estimators.

Expanding on this approach, \citet{Lam2018} employed deep neural 
networks to predict the stability of initially coplanar, circular P-type orbits 
for circumbinary planets. Their method achieved an accuracy of at least 
$86\%$ for the test set,
further demonstrating the potential of ML in analyzing orbital
stability.

Building upon the successes of previous ML
applications in orbital stability analysis, 
we propose using this approach to predict a stable region map 
surrounding a single-planet system. 
This method can also be extended to analyze the stability of other analogous systems, 
such as planet-satellite pairs, planetary rings, and binary minor planets, 
provided they lie within the trained parameter space of the ML model.

We conducted a series of dimensionless numerical 
simulations involving an elliptical three-body problem 
(star, planet, and a particle orbiting the planet).
Each simulation represents a distinct initial condition 
within our dataset, used to train and evaluate machine 
learning algorithms. 
Expanding on the work of \citet{Domingos2006}, 
our study explores a wide range of orbital and physical 
parameters, including the mass ratio of the system and
several orbital elements of both the particle and planet. 

This comprehensive approach could be 
computationally demanding if solved purely through 
numerical simulations, especially when increasing 
the number of grid parameters to be analyzed. 
Running a single three-body problem may be relatively fast, 
but a broader set of initial conditions may significantly 
increases computational costs, highlights the 
advantages of our ML method.

This paper is structured as follows:
Section \ref{S-SupervisedLearning} introduces key concepts and methodologies for our analysis
and introduce ML algorithms used in this study.
Section \ref{S-numericalsumulations} presents a description of
our numerical model and dataset preparation. Section \ref{S-imbalancedclass} describes how to tackle with imbalanced classes.
Section \ref{S-PerformanceMeasures} presents the performance evaluation of our
ML algorithms.
Section \ref{S-application} compares the predictions made by our
best performance ML model with the results from \citet{Domingos2006} and
\citet{Pinheiro2021}, as well as with a stability map of Saturn's
satellites. Finally, Section \ref{s-finalremarks} presents our concluding remarks and
implications of this study.


\section{Supervised learning \label{S-SupervisedLearning}}

This section presents fundamental concepts and
techniques for our study.
These definitions and approaches will be referenced and
compared throughout the subsequent sections, particularly in our
results and discussion.

The Supervised learning includes 
in the training data the desired solution (labels). 
These labels can be composed of continuous numerical 
value (regression task) or discrete values 
which represent different classes (classification task).

Classification tasks are techniques that use 
training samples (or instances) during the learning process
to develop a generalization of classes and
make a prediction given unknown new instances. 
Here, we address binary classifications 
stable and unstable particles to 
predict stability maps within a specific planetary system.

The dataset was split into three subsets: training, validation, and test data.
The training data is used by the algorithm to find the best generalization
features for different classes. The validation data provides an unbiased
evaluation of model effectiveness, and is used to tune model hyperparameters
and thresholds, select the best 
ML algorithm, and detect
underfitting and overfitting \citep{Shalev2014}.

The cross-validation technique involves partitioning the training data into
subsets. Specifically, in this paper, we split the training data into 5
subsets. Subsequently, we train and evaluate the model on distinct subsets,
followed by averaging the outcomes to estimate the
unbiased error on the training set of the model
\citep{Fushiki2011}.

After training several models with different configurations multiple times on
the reduced training set and evaluating them through cross-validation, we
identified the best-performing model. Subsequently, we retrained the selected
model using the complete training dataset and evaluated 
final performance of ML model
using holdout data (test dataset).

\begin{figure}[ht]
    \centering
    \resizebox{0.45\textwidth}{!}{\includegraphics{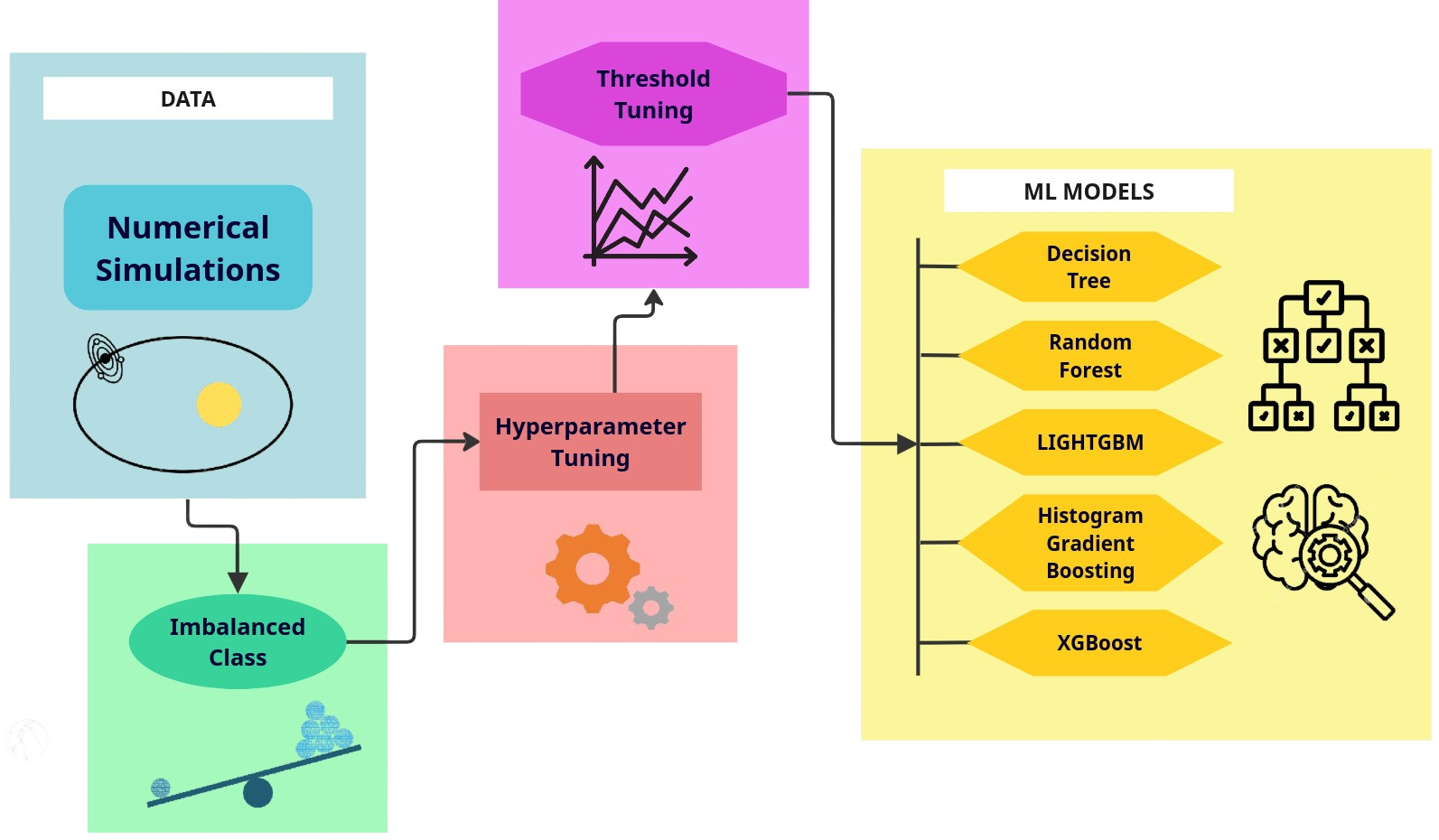}}
\caption{Flowchart illustrating the workflow of this paper.}
\label{WorkFlows}
\end{figure}

We divided our work into five main steps, as shown in Figure~\ref{WorkFlows},
which sketches the workflow of our methodology. The process
begins with generating a dataset from numerical simulations. We then address
the issue of imbalanced classes through resampling techniques. Using
cross-validation, we perform hyperparameter tuning and explore optimal
threshold values to establish the best configuration model. The final step
involves assessing the performance of the best model. 

\subsection{Hyperparameter tuning}

Hyperparameters are specific algorithm parameters that must be set prior to the
training process, and the model performance directly depends on them
\citep{Weerts2020}. The search for the best hyperparameters was carried
out in two different ways: Grid and Random Search, both validated using
cross-validation.

The Grid Search systematically explores all possible combinations of
hyperparameters. In contrast, Random Search randomly selects some
combinations to evaluate, which is useful when dealing with an extensive
number of potential combinations.

\subsection{Threshold tuning}
In general, in binary classification tasks, classifier algorithms employ
a default threshold value of 0.5. This means that if the probability of an
instance being in a positive class is $\geq$ 0.5, it is assigned to the
positive class; otherwise, it is classified as negative. However, for an
imbalanced dataset, using 0.5 as the threshold value is generally
inappropriate \citep{Zou2016}.

\begin{figure}[ht]
    \centering
    \resizebox{\hsize}{!}{\includegraphics{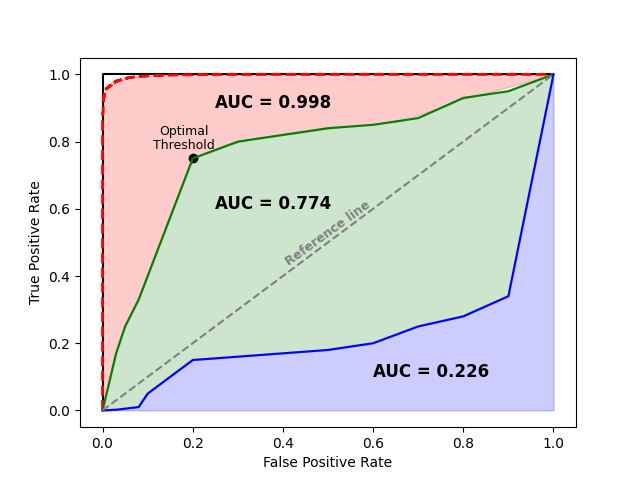}}
    \caption{The ROC curve representation of three hypothetical models. 
    The black curve behaves like an ideal model,
    while the green and blue curves represent good
    and bad classifiers, respectively.
    The gray dashed line is the reference line.
    The red curve represents 
    the performance of our best-performing ML model.}
    \label{threshold}
\end{figure}

The Receiver Operating Characteristic (ROC) curve is a frequently employed tool
for determining the optimal threshold value \citep{Geron2022hands}. The ROC
curve is a probability curve, which plots the True Positive Rate (hereafter
referred to as recall) versus the False Positive Rate for different
threshold probabilities from 0 to 1. Figure \ref{threshold} provides an
illustrative example of ROC curves for three hypothetical models 
with varying performance levels, and also the 
ROC curve of our best performance ML model (see sec.~\ref{S-PerformanceMeasures}). 

The True Positive Rate is the ratio of positive instances that are
correctly predicted, while the False Positive Rate represents the fraction of
negative instances that are incorrectly predicted as positive.
The ROC curve always begins at the point (0,0), where the threshold is 1,
signifying that all instances are predicted as the negative class. It ends at
(1,1), where the threshold is 0, indicating that all instances are predicted as
the positive class.

The Area Under the Curve (AUC) represents the area under the ROC curve,
providing a measure of the model's ability to differentiate classes. In a
perfect model (exemplified by the black curve in figure \ref{threshold}), the
positive class is always predicted with $100\%$ accuracy regardless of the
threshold parameter, resulting in an AUC of 1. 

Good predictors tend to have an AUC value close to $1$. For instance,
the green curve in figure \ref{threshold}
has an AUC of $0.77$, indicating that the model has a $77\%$ probability of
successfully distinguishing between positive and negative classes. 
The dashed red line represents the ROC curve of our 
best-performing ML model with an AUC of 0.998. 
This model will be discussed in section \ref{S-PerformanceMeasures}.

The reference line (dashed gray line) represents an AUC of 0.5, signifying that the
model cannot differentiate between classes, resulting in random guesses for
predictions. Conversely, the blue curve represents a poorly performing model
with an AUC of 0.23, where the majority of positive instances are predicted as
negative.

The optimal threshold (the black dot in Figure \ref{threshold})
can be determined as the point closest to the upper-left
corner of the ROC curve (0,1),
or the point that maximizes the distance
from the reference line
where the true positive rate is high and
the false positive rate is low.

The concepts and methods outlined in this section provide the framework for
our subsequent analysis. We will refer back to these definitions and compare
their effectiveness in later sections, particularly when evaluating our
results and discussing their implications.

\subsection{Classification algorithms}
\label{S-ClassificationAlgorithms}

We tested eight distinct ML algorithms 
to find the optimal approach: 
Nearest Neighbors, Naive Bayes, Artificial Neural Network, and tree-based methods
Decision Tree, Random Forest, 
XGBoost, Light Gradient Boosting Machine (LightGBM), and 
Histogram Gradient Boosting. Lately, we ran a Genetic algorithm to verify our findings.

Given that all tree-based algorithms demonstrated
better and comparable performance, we will focus our analysis on these methods. 
With the exception of Decision Tree, 
all other tree-based classifiers are ensemble methods, 
which utilize a group of weaker learners 
(in this case, an ensemble of decision trees) to enhance their predictions. 
The collective perspective of a group of predictors yields better predictions than a
single predictor \citep{Geron2022hands}. 
We give further details of
the fundamental concepts of each of these 
tree-based algorithms in Appendix \ref{Ss-Algorithms}.


\section{Dataset preparation
\label{S-numericalsumulations}}

In this section, we detail the methodology for investigating the orbital
stability of a three-body problem (star, planet, and particle) through an
ensemble of dimensionless numerical simulations. Each simulation represents a
distinct initial condition for a hypothetical planetary system.

We begin by defining the mass parameter $\mu$ as 

\begin{align}
\mu = \frac{m_\mathrm{p}}{M_\mathrm{s} + m_\mathrm{p}},
\label{mass_ratio}
\end{align}

\noindent where $m_\mathrm{p}$ and $M_\mathrm{s}$ are the
masses of the planet and star, respectively.

A survey including 4,150 confirmed exoplanets 
from \citet{Schneider2011}, 
reveals that 95.71 $\%$ of them 
have an equivalent $\mu$ $\leq$ $10^{-2}$, 
and 2.05 \% of these have $\mu$ < $10^{-5}$. 
Thus, we restricted our analysis to this range of 
mass parameter, and within it we 
adopted  a random uniform distribution.

Throughout all numerical simulations,
the star and planet masses were set 
by $M_\mathrm{s} = 1 - \mu$ and
$m_{p} = \mu$, respectively.
The planet semi-major axis was set to $1$ 
and its collision radius ($r_\mathrm{c}$) was
defined as  $5 \%$ of the Hill radius ($R_{\textrm{Hill}}$) 
computed at the pericentre:  

\begin{align}
    r_{c} =  0.05(1 - e_{p} )\sqrt[3]{\frac{\mu}{3}}.
    \label{Hill}
\end{align}

\begin{figure}[ht]
    \centering
    \resizebox{0.45\textwidth}{!}{\includegraphics{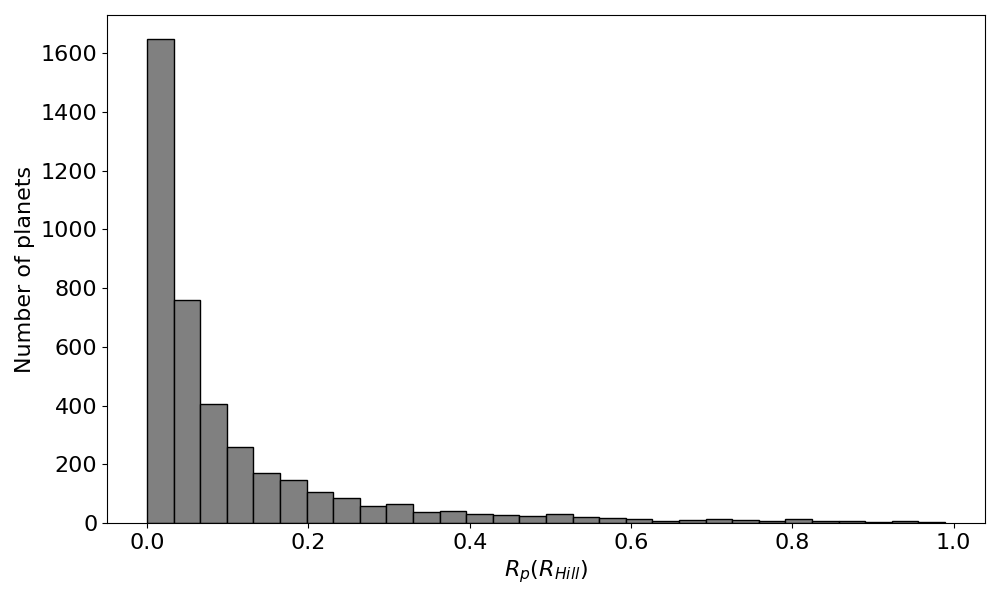}}
    \caption{Histogram of the number of confirmed planets in units of
    their Hill radius from a survey consisting 
    4,150 different planets \citep{Schneider2011}.}
    \label{histogram_mass}
\end{figure}

This choice for $r_c$ represents
55.6\% of planets in the catalogue of \citet{Schneider2011} 
have a collision radius that is less than 5\% of their Hill radii,
as shown in the histogram in figure \ref{histogram_mass}.

The planet eccentricity ($e_\mathrm{p}$) and true anomaly ($f_\mathrm{p}$)
were randomly uniformly sampled within the specified ranges:
$e_{p}~=~[0-0.99]$ and $f_{p} =  [0^\circ - 360^\circ]$.
The remaining orbital elements for the planet
inclination ($i_\mathrm{p}$), arguments of pericenter ($\varpi_\mathrm{p}$), 
and longitude of the node ($\Omega_\mathrm{p}$) were set to 
zero.

The test particle was modeled orbiting a
planet and gravitationally disturbed by the star.
We adopted a random uniform distribution to select
particle orbital initial conditions for semi-major axis 
from 1.1 $r_{c}$ to 1 $R_{\textrm{Hill}}$, 
eccentricity from 0 to 0.99, 
inclination between $0^\circ$ and $180^\circ$, and 
arguments of pericenter, longitude of the node, and true anomaly
ranging from $0^\circ$ to $360^\circ$.

We numerically integrated using \textsc{Rebound} package
and IAS15 integrator \citep{Rein2015},
for a time span of $10^4$ orbital period of the planet.
We recorded every time when particle collided with the planet 
or was ejected from the system, which occurred when 
the particle reaches a hyperbolic orbit (eccentricity $e \geq 1$)
or when its semi-major axis is $a > 1$.

In total, we ran $10^{5}$ numerical systems
and the overall outcome was as follows: 
$11.83\%$ of the particles remained stable 
throughout the entire integration,
while $88.17\%$ were unstable. 
Among the unstable particles, 
53.56\% of them collided with the planet, 
and 46.44\% were ejected from the system.

The outcomes of the numerical simulations were utilized 
to build a dataset for training ML algorithms.
This dataset consisted of 9 features, 
which are the initial conditions of each numerical simulation:
the mass ratio of the system, 
semi-major axis, inclination, argument of pericenter, 
and longitude of the node of the particle, 
and eccentricity and true anomaly of both the planet and the particle.
We labeled each initial condition with the numerical 
results as either stable or un unstable system.

\begin{figure}[ht]
    \resizebox{\hsize}{!}{\includegraphics{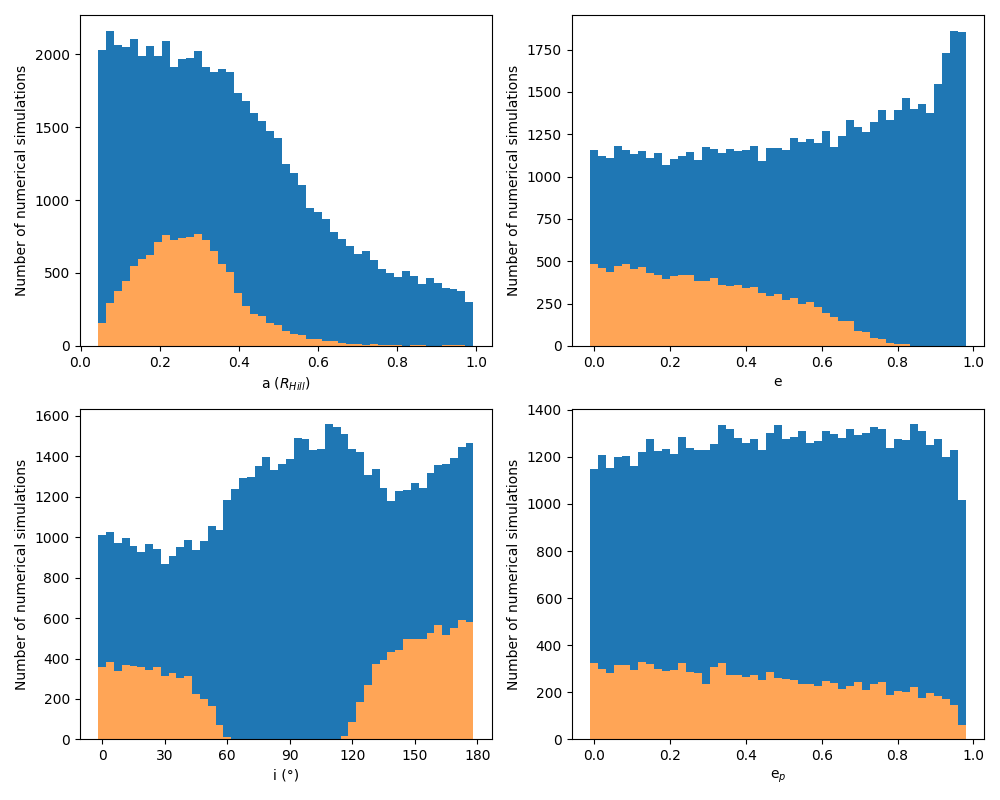}}
    \caption{Histogram of the distribution of stable (orange) 
    and unstable (blue) systems across the range of
    particle semi-major axis (upper left panel) and particle eccentricity (upper right panel)
    particle inclination (lower left panel and planet eccentricity (lower right panel).}
    \label{Histogram}
\end{figure}

Figure \ref{Histogram} shows four histograms
of the number of numerical simulations in relation to the
particle semi-major axis (upper left panel),
eccentricity (upper right panel),
inclination (lower left panel),
and planet eccentricity (lower right panel).

As the eccentricity of the planet increases, the number 
of stable systems is slightly reduced. Nevertheless, we can find 
stable particles even for highly eccentric planets. 
The particle eccentricity has a more distinct effect on the number of stable systems, 
with no stable orbits found for  $e > 0.8$. This is expected since 
particle orbits experience stronger 
gravitational disturbances at their pericenters, which
increases the quantity of particles that collided. 

The number of stable systems is also strongly 
influenced by the inclination of particles 
relative to the orbital plane of the planet.
In our set, there are no stable systems for inclination between $80^\circ$ and 
$110^\circ$, while retrograde systems tend to have more stable orbits compared to 
prograde ones.

Closer to the planet there is a 
significant number of collisions, and 
the number of stable systems increases with the initial particle distance,
reaching a peak at $0.3~\rhill$. 
After this peak, the quantity of stable systems declines, and
only a few systems remain stable beyond $0.7~\rhill$,
with the majority of particles being ejected.

Other features, such as the mass ratio of the system, 
argument of pericenter, longitude of the ascending node, and 
true anomaly, showed a regular distribution
across the range of their parameters.

\section{Imbalanced class
\label{S-imbalancedclass}}

The numerical results demonstrate that the classes are
disproportionate in the dataset by a factor 1:10, with the unstable 
class being much larger than the stable class. 
This imbalance could lead to biases in classification and impair
the effectiveness of the various classifier algorithms
when trying to generalize \citep{Zheng2015}.
ML algorithms typically assume
that classes are evenly distributed,
which can causes the minority class to be
classified poorly \citep{Kumar2012,Carruba2020}.

Resampling the training data is one strategy that
can be used to tackle this problem.
One option is the undersampling, which is a technique that removes instances
from the majority class.
The weakness of this method is that it may lose
information by removing part of the data \citep{Mohammed2020}.
Another resampling process is the oversampling,
which involves increasing the size of
the minority class until the classes are balanced.
We tested the performance of four
different types of oversampling: random oversampling,
Synthetic Minority Over-sampling Technique
(SMOTE), Borderline SMOTE, and  Adaptive Synthetic Sampling (ADASYN).
In Appendix \ref{S-Imbalaced}, we provide additional details on the 
key concepts of each of these resampling techniques.

According to \cite{Carruba2023}, an severe imbalanced dataset
can occur when the class ratio is 1:100, and in this scenario  resampling methods may be necessary. 
After testing the performance of our algorithms with various resampling techniques, 
we observed a marginal improvement, except for the LightGBM that increased of 13\% in identifying stable particles,
achieved solely through the resampling technique, without hyperparameter or threshold tuning.  
With this improvement, the LightGBM performance 
is almost as good as the 
best-performing model without any resampling.

\section{Performance evaluation}\label{S-PerformanceMeasures}

Evaluation metrics enable the assessment of classifier performance.
The confusion matrix is a statistical table that maps prediction
results, showing the quantity of correctly and incorrectly predicted data
\citep{Stehman1997}. While accuracy measures the rate at which a model
correctly predicts, this metric may not be reliable for imbalanced datasets.
A confusion matrix displays the actual number of instances for each class
in its rows, and the predicted number of instances for each class in
its columns. It is divided into four categories: True Positive (TP) and
True Negative (TN), where the algorithm correctly predicts positive and
negative classes; and False Positive (FP) and False Negative (FN),
where the algorithm incorrectly predicts positive and negative classes.

These four categories are used to measure recall, specificity, precision,
F1-score, and accuracy. Accuracy is the ratio of instances correctly
predicted by the algorithm, given by
\begin{align}
\text{Accuracy} = \frac{\text{TP} + \text{TN}}{\text{TP} + \text{TN} + \text{FP} + \text{FN}}
\label{accuracy}
\end{align}

Specificity or true negative rate is the ratio of correctly predicted
instances within the actual negative class:
\begin{align}
\text{Specificity} = \frac{\text{TN}}{\text{TN} + \text{FN}}
\label{specifity}
\end{align}
\noindent and the False Positive rate is given by $1 - \text{Specificity}$.

{Recall or the true positive rate is the ratio of correctly classified positive instances 
to the total number of actual positive instances:
\begin{align}
\text{Recall} = \frac{\text{TP}}{\text{TP} + \text{FN}}
\label{specifity}
\end{align}
}

Precision is the ratio of actual positive instances within the
predicted positive class:

\begin{align}
\text{Precision} = \frac{\text{TP}}{\text{TP} + \text{FP}}
\label{precision}
\end{align}

The F1-score combines precision and recall by calculating their harmonic mean:
\begin{align}
\text{F1-score} = 2~\left(\frac{\text{Precision} \times \text{Recall}}{\text{Precision} + \text{Recall}}\right)
\label{f1}
\end{align}

While accuracy measures the rate at which a model correctly predicts, this
metric may not be reliable for imbalanced datasets. F1 scores measure the
overall quality of a model, ranging from 0 to 1, where 1 signifies a model
that perfectly classifies instances and 0 indicates a model that fails to
classify any instance correctly.

\subsection{Results and Comparison of Different Algorithms} \label{Ss-PerformanceMLmodels}

The ML outcomes were evaluated using a dataset generated from
numerical simulations (see Sec. \ref{S-numericalsumulations}). 
Our binary approach involves classifying systems as either
stable or unstable. The final dataset consists of 100,000 simulations, divided into
three subsets: 80\% for training and validation data and 20\% for testing data.
In the presence of imbalanced classes and to achieve better performance in
classifier algorithms, we tested five distinct resampling methods. 
We utilized
cross-validation techniques to identify the best hyperparameters, resampling
method, and the optimal threshold value for each algorithm.

The best performance among the five algorithms was achieved by XGBoost, without
using any resampling techniques, with a threshold value of 0.4, and setting some
hyperparameters as follows:

\begin{enumerate}[i.]
\item booster = gbtree.Booster is a hyperparameter for
the type of boosting model used during the training stage. Gbtree
means gradient boosted trees.
\item eta = 0.2. Eta is the learning rate and mathematically,
it scales the contribution of each tree added to the model.
\item scale\_pos\_weight = 0.4. Controls the balance of positive
and negative class weights.
\item max\_depth = 8. The maximum depth for each tree.
\end{enumerate}

This result was also verified using the Genetic Algorithm. Genetic Algorithms
mimic the process of genetic evolution and are capable of selecting the most
optimal algorithm and the best hyperparameters for a given dataset
\citep{Chen2004}.

\begin{figure}[!h]
\includegraphics[width=0.49\linewidth]{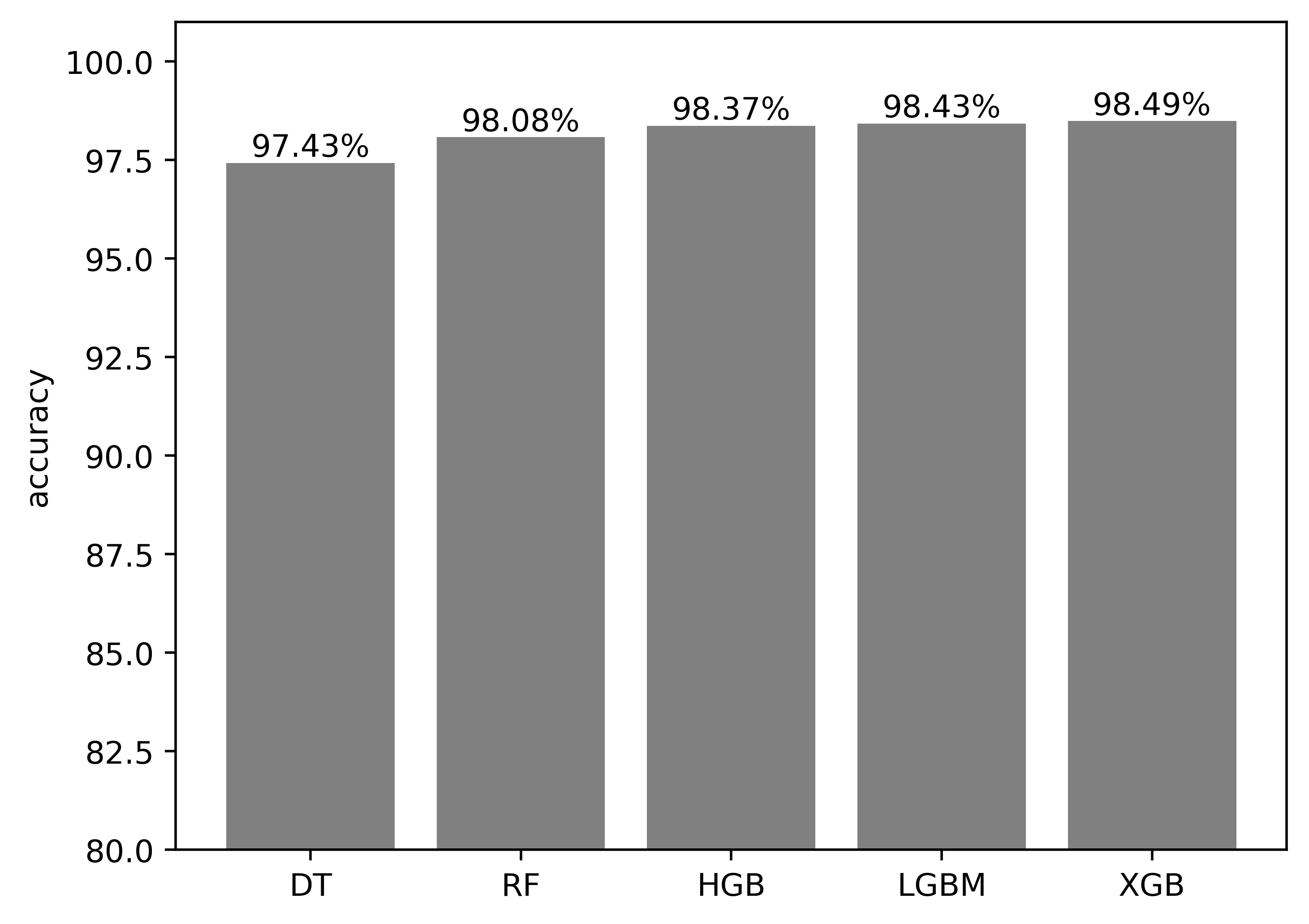}
\hfill
\includegraphics[width=0.49\linewidth]{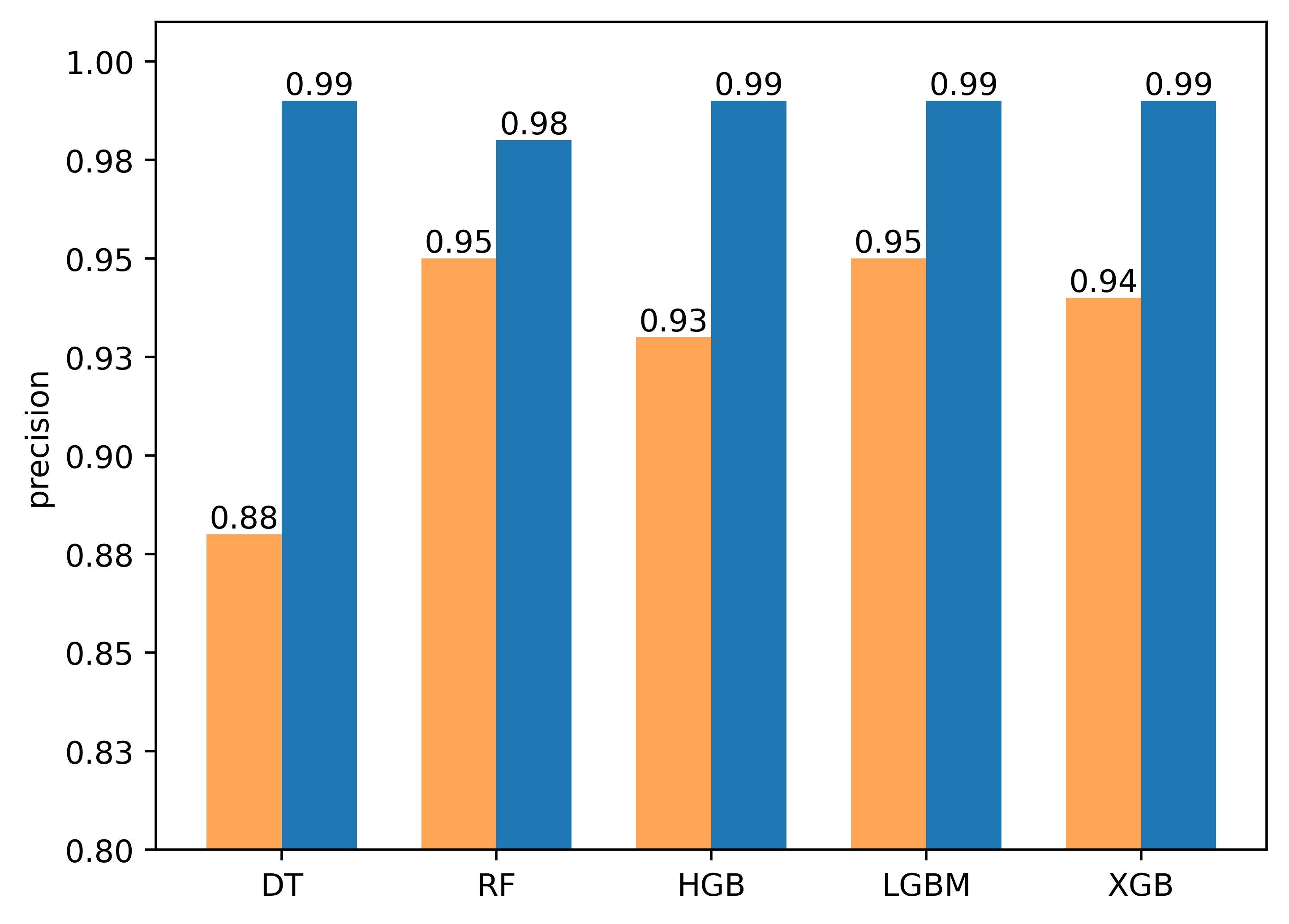} \\
\includegraphics[width=0.49\linewidth]{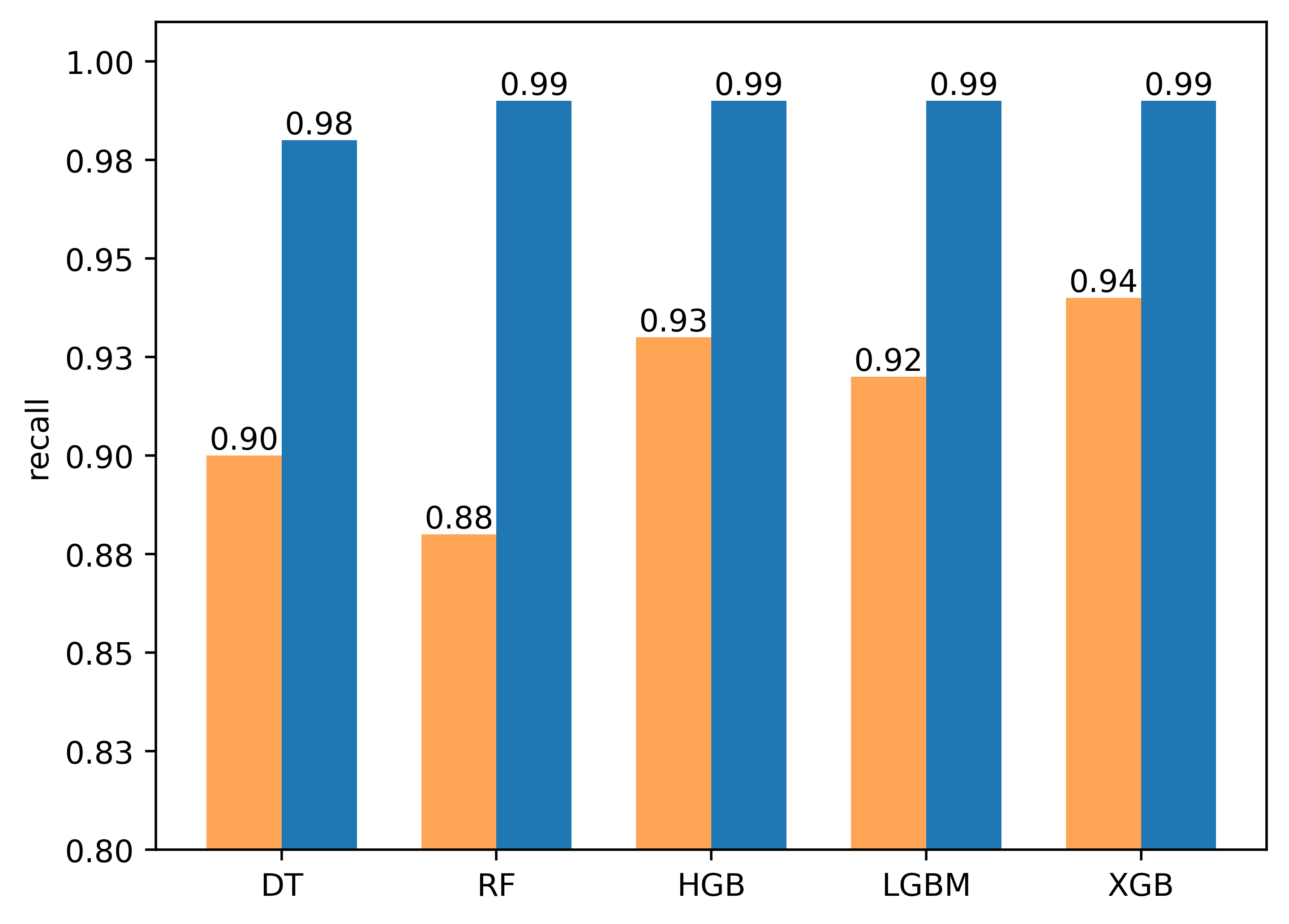}
\hfill
\includegraphics[width=0.49\linewidth]{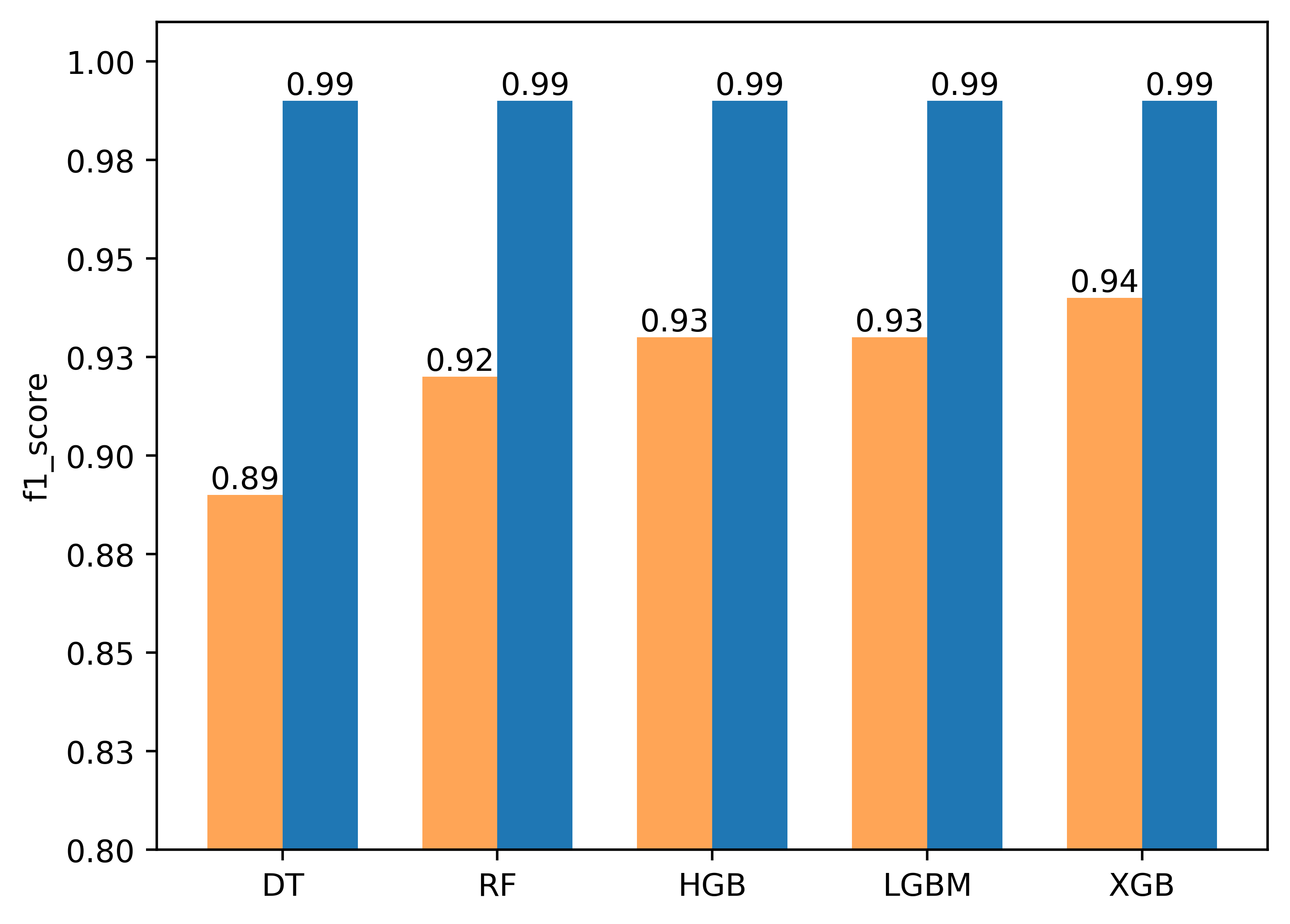}
\caption{
The performance of the best model among the five tested algorithms.
The upper panels show accuracy and precision, while the lower panels
show recall and F1-score. The stable and unstable classes are represented
by orange and blue bars, respectively. 
The algorithms are abbreviated as follows:
Decision Tree (DT), Random Forest (RF), Histogram Gradient Boosting (HGB), Light Gradient Boosting Machine (LGBM), 
Extreme Gradient Boosting (XGB).}
\label{metrics}
\end{figure}

Figure \ref{metrics} shows the performance of the five algorithms using four
distinct metrics. The upper left panel displays accuracy, while precision for each
class is presented in the upper right panel. The lower left panel shows the
recall, and the lower right panel shows F1-scores. Stable classes are
represented by {orange} bars, while unstable classes are 
shown with {blue} bars.

All algorithms demonstrated comparable accuracy performance.
XGBoost achieved the highest accuracy at 98.48\%, while Decision Tree exhibited
the lowest score at 97.43\%, showing a marginal difference of 1.05\%.

An important consideration here is that accuracy measures how correctly the model
classifies whether the system is stable or unstable across all predicted data.
Therefore, this favorable result may be influenced by the proportion of
imbalanced classes, with the quantity of unstable instances being
approximately 7.5 times larger than stable ones in our dataset. The algorithms
achieved a recall and precision of 99\% for all classes, except for the
precision of Random Forest and the recall of the Decision Tree
algorithm, which were 98\%.

In terms of stable class performance, both LightGBM and Random Forest achieve the
highest precision value, correctly classifying 95\% of their predictions. They
identify 92\% and 88\% of actual stable instances (recall), respectively. XGBoost
attains the same percentages for precision and recall, leading to the highest
F1-score of 94\%, while LightGBM and Random Forest achieve 93\% and 92\%,
respectively.

The area under the ROC curve (AUC) provides a measure of overall model performance,
with a value of 1 representing perfect classification (Figure \ref{threshold}).
Table \ref{AUCvalues} compares the AUC values for each algorithm. XGBoost
outperforms other models, closely followed by LightGBM and Histogram Gradient Boosting. 
The Decision Tree algorithm shows slightly lower
performance compared to the ensemble methods.

\begin{table}[ht]
    \caption{Area under the ROC Curve for the best model of each algorithm, 
    sorted in descending order of performance.}
    \label{AUCvalues}
    \centering
    \begin{tabular}{cc}
    \hline
         Model & AUC \\
    \hline
    XGBoost & 0.9978\\
    LightGBM &  0.9975\\
    Histogram Gradient Boosting & 0.9974\\
    Random Forest & 0.9957\\
    Decision Tree &0.9554\\
    \hline
    \end{tabular}
\end{table}

\begin{figure}[!h]
\centering
\begin{subfigure}{0.49\linewidth}
  \includegraphics[width=\linewidth]{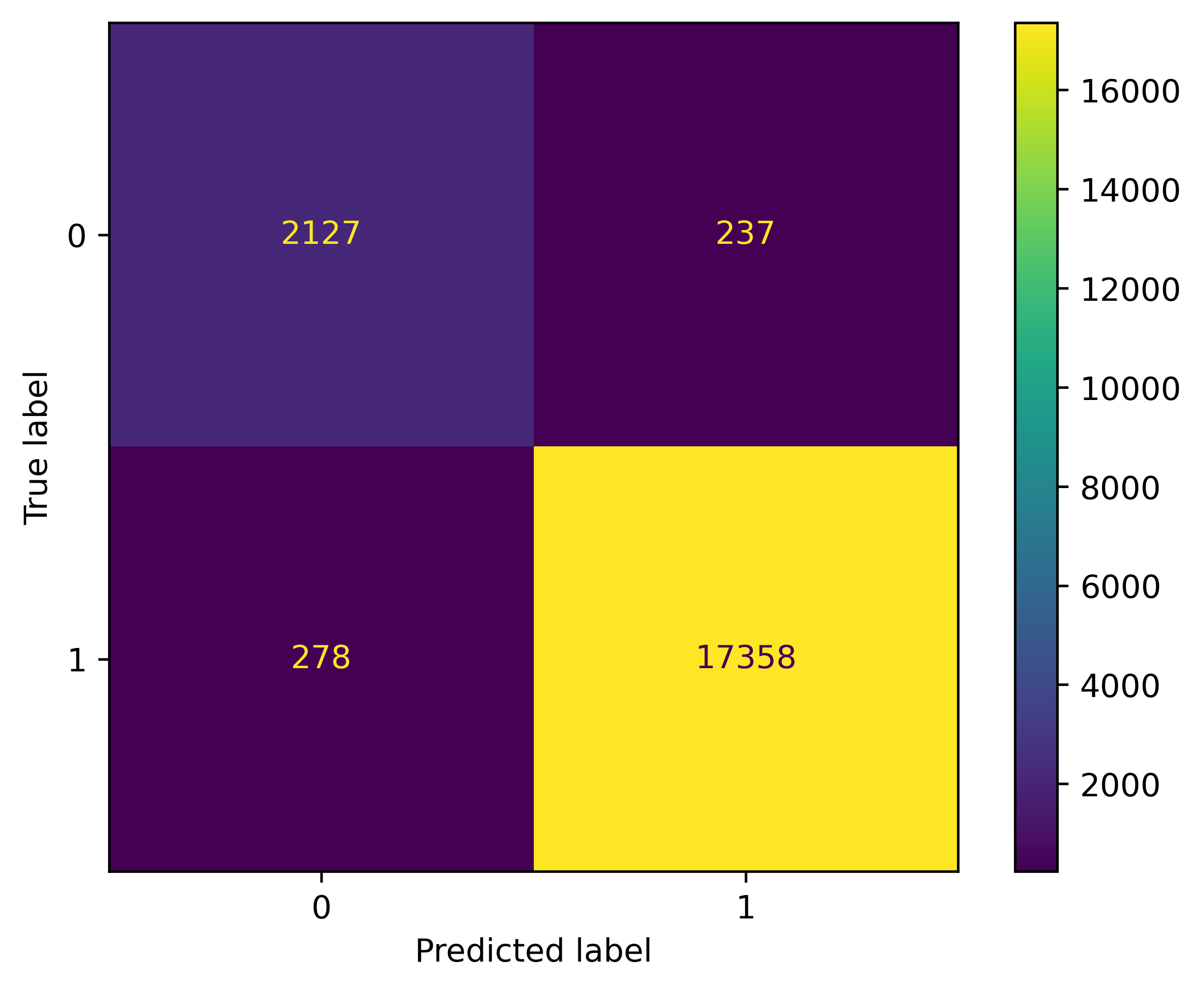}
  \caption{Decision Tree}
\end{subfigure}
\hfill
\begin{subfigure}{0.49\linewidth}
  \includegraphics[width=\linewidth]{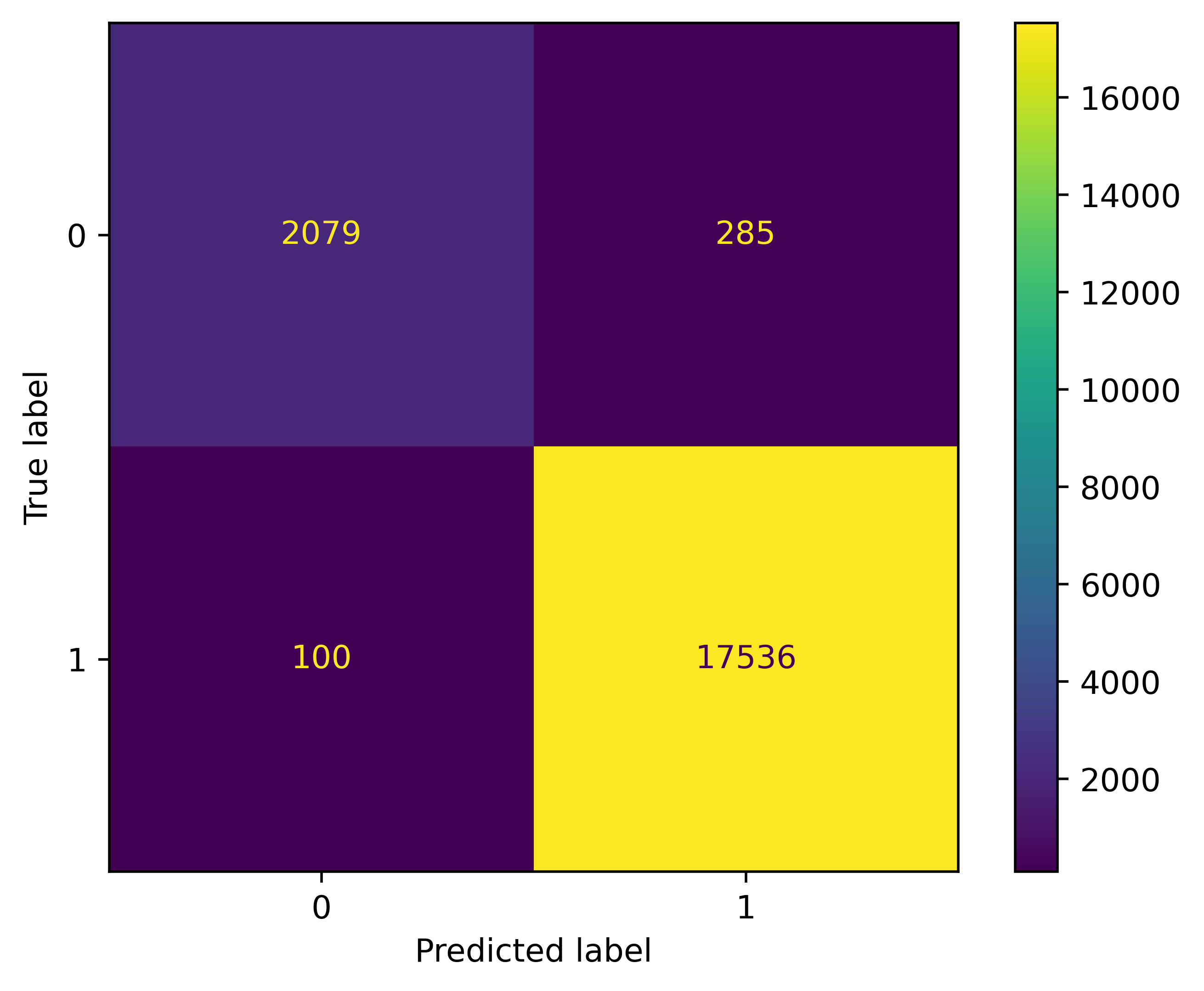}
  \caption{Random Forest}
\end{subfigure}
\hfill
\begin{subfigure}{0.49\linewidth}
  \includegraphics[width=\linewidth]{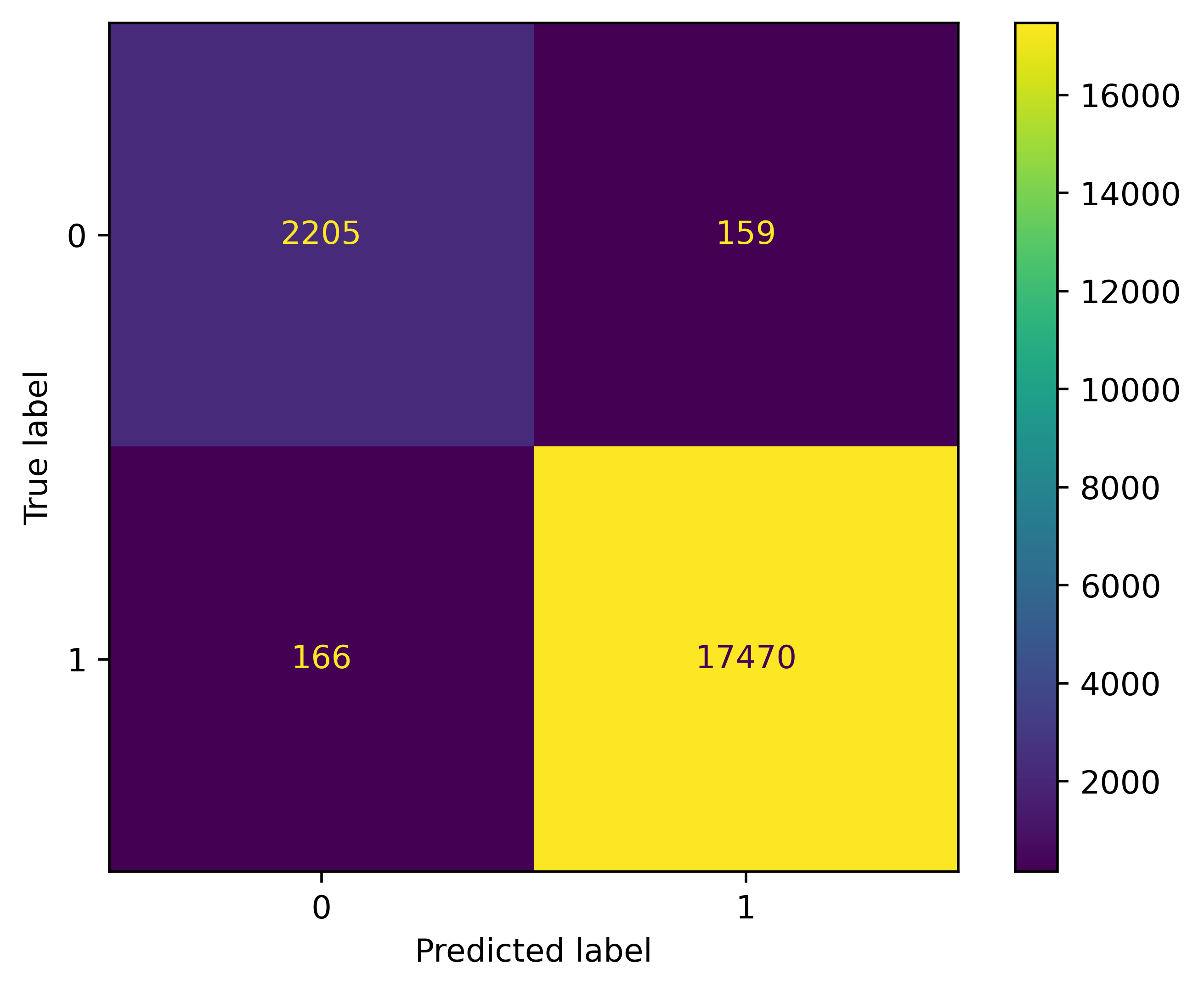}
  \caption{Histogram Gradient Boosting}
\end{subfigure}
\hfill
\begin{subfigure}{0.49\linewidth}
  \includegraphics[width=\linewidth]{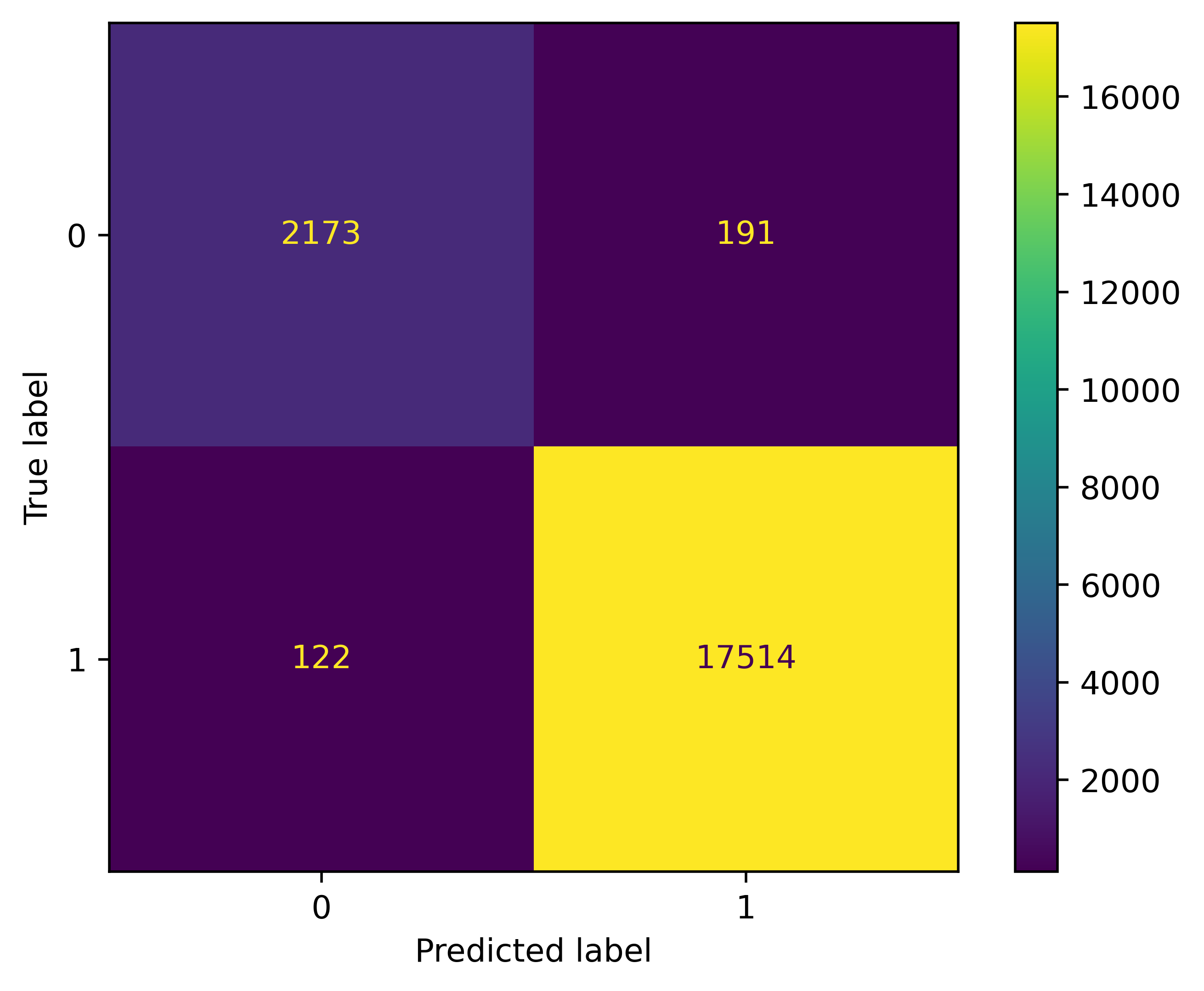}
  \caption{LightGBM}
\end{subfigure}
\hfill
\begin{subfigure}{0.49\linewidth}
  \includegraphics[width=\linewidth]{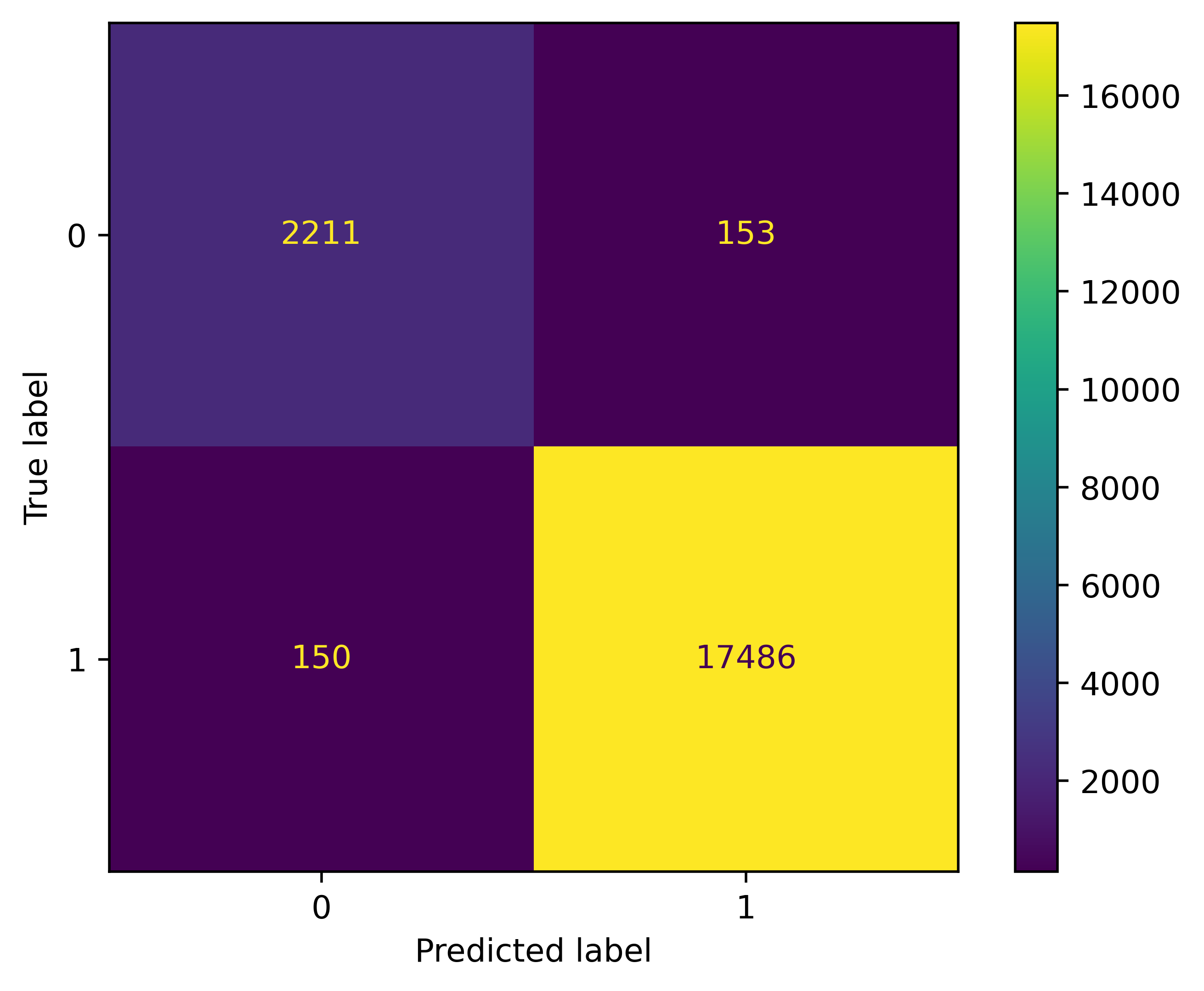}
  \caption{XGBoost}
\end{subfigure}
\caption{Confusion matrices for the best models of five
ML algorithms.}
\label{confusionmatrix}
\end{figure}

Figure \ref{confusionmatrix} shows the confusion matrices for each algorithm,
providing statistical results from the testing dataset of 20,000 instances.
Class 0 represents a stable system, while class 1 is unstable. All algorithms
demonstrate satisfactory performance, with even the lowest-performing model
(Decision Tree) correctly predicting at least 19,485 instances.

XGBoost, the highest accuracy algorithm, misclassified only 303 instances,
with a nearly balanced distribution of errors between stable and unstable classes 
(150 and 153 instances, respectively). 
It also correctly identified the highest number of stable instances (2,211).
Random Forest and LightGBM achieved the highest precision for the stable class
(0.95), slightly outperforming XGBoost in this metric.

Random Forest showed the best performance in correctly classifying unstable
instances (17,536). However, it also had the highest number of false positives,
misclassifying 285 stable instances as unstable – nearly double the error rate
of XGBoost for this class.

\subsection{Feature importance} \label{Ss-featureimportance}

\begin{figure}[ht]
    \resizebox{0.47\textwidth}{!}{\includegraphics{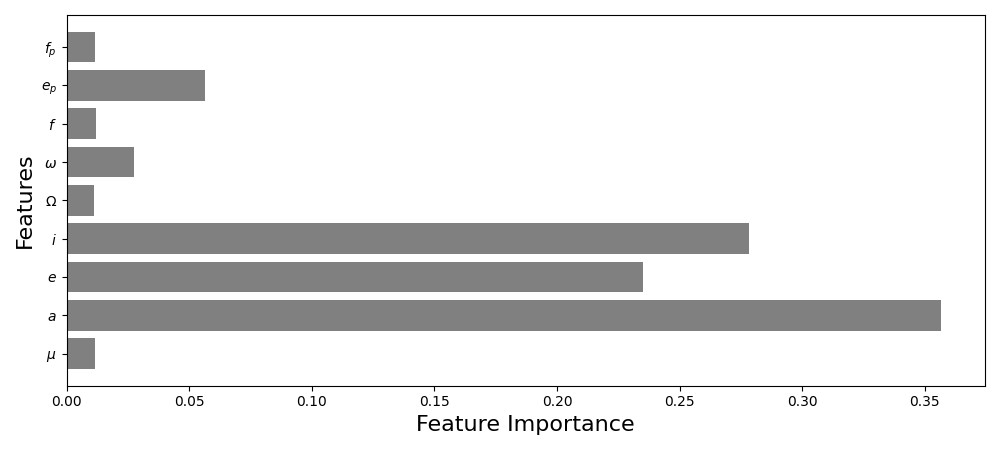}}
    \caption{Feature importance derived from the XGBoost model. Orbital elements
are denoted as follows: semi-major axis ($a$), eccentricity ($e$),
inclination ($i$), argument of pericenter ($\omega$), longitude of node
($\Omega$), and true anomaly ($f$). Planet-specific parameters are
indicated with a 'p' subscript. The mass ratio of the system is represented
by $\mu$.}
    \label{f-featureimportance}
\end{figure}

An interesting outcome from the XGBoost algorithm is the feature importance,
which estimates how effectively each feature contributes to reducing impurity
throughout all the decision tree splits. The feature importance is determined by
averaging the importance of each feature across all decision trees, 
and Figure \ref{f-featureimportance} illustrates the
relative importance of features in our classification task.

Particle orbital elements dominate the top three positions, with semi-major
axis ($a$), eccentricity ($e$), and inclination ($i$) scoring 0.3567, 0.2783,
and 0.2351, respectively. These findings align with the distribution patterns
observed in Section \ref{S-numericalsumulations}, where we noted significant
reductions in stable systems beyond $0.6 r_\mathrm{Hill}$ and the absence of
stable particles with inclinations near $90^\circ$ or eccentricities $\geq 0.8$.
Planet eccentricity ranks fourth in importance (0.0564), followed by particle
argument of pericenter ($\omega$, 0.0274). The remaining features - particle
true anomaly ($f$), mass ratio ($\mu$), planet true anomaly ($f_\mathrm{p}$),
and particle longitude of node ($\Omega$) - contribute less significantly,
with scores ranging from 0.0119 to 0.0110.

These results quantify the relative importance of orbital parameters in
determining system stability, that most significantly influence dynamical outcomes.


\section{Comparative analysis}
\label{S-application}
We now apply our best ML model, XGBoost, to predict four
distinct diagrams of initial conditions ($a \times e$) with the aim of
reproducing the results of \cite{Domingos2006}, \cite{Pinheiro2021}, and those
for a particular Saturn moon group known as Inuit. 
In each diagram  the test particles are uniformly distributed within
a range of initial semi-major axes from 1.1 $r_\mathrm{p}$ to 1 $r_\mathrm{Hill}$
with a step size of $\Delta a = 0.1 r_\mathrm{p}$, and initial eccentricities
ranging from 0 to 0.5, with $\Delta e = 0.01$.

The XGBoost model, which demonstrated superior performance in our
earlier analysis with an accuracy of 98.48\%, is now tasked with predicting
stability in these specific planetary systems. This approach allows us to
evaluate the model's generalization capabilities across diverse orbital
configurations and compare its predictions directly with established numerical
results.

To create an entire stability map with 
$\sim$ 10,000 different initial conditions, 
numerical simulations required between two and 
four days on a single core of an Intel i7-1165G7 processor, 
while ML predictions, with saved training, generate a map in 0.5 seconds.
\subsection{Comparison with \citet{Domingos2006}}

\cite{Domingos2006} numerically simulated multiple stable maps
to delineate the boundaries of stable regions surrounding
a close exoplanet orbiting its star at a semi-major axis of 0.1 au.
They derived an analytical expression for the critical exosatellite
semi-major axis as a function of the eccentricity of both the satellite and
the planet, for two different orbital inclinations $0^\circ$ and $180^\circ$.

We reproduce ($a \times e$) diagrams of two different systems explored
by \cite{Domingos2006}, one in prograde and the other in retrograde orbit.
The mass ratio between planet and star was defined as $\mu = 10^{-3}$,
with a planet radius $r_\mathrm{p} = 0.05 r_\mathrm{Hill}$ and orbiting
a Sun-like star in a circular orbit.

\begin{figure*}[h]
  \centering
  \begin{subfigure}{0.49\linewidth}
    \includegraphics[width=\linewidth]{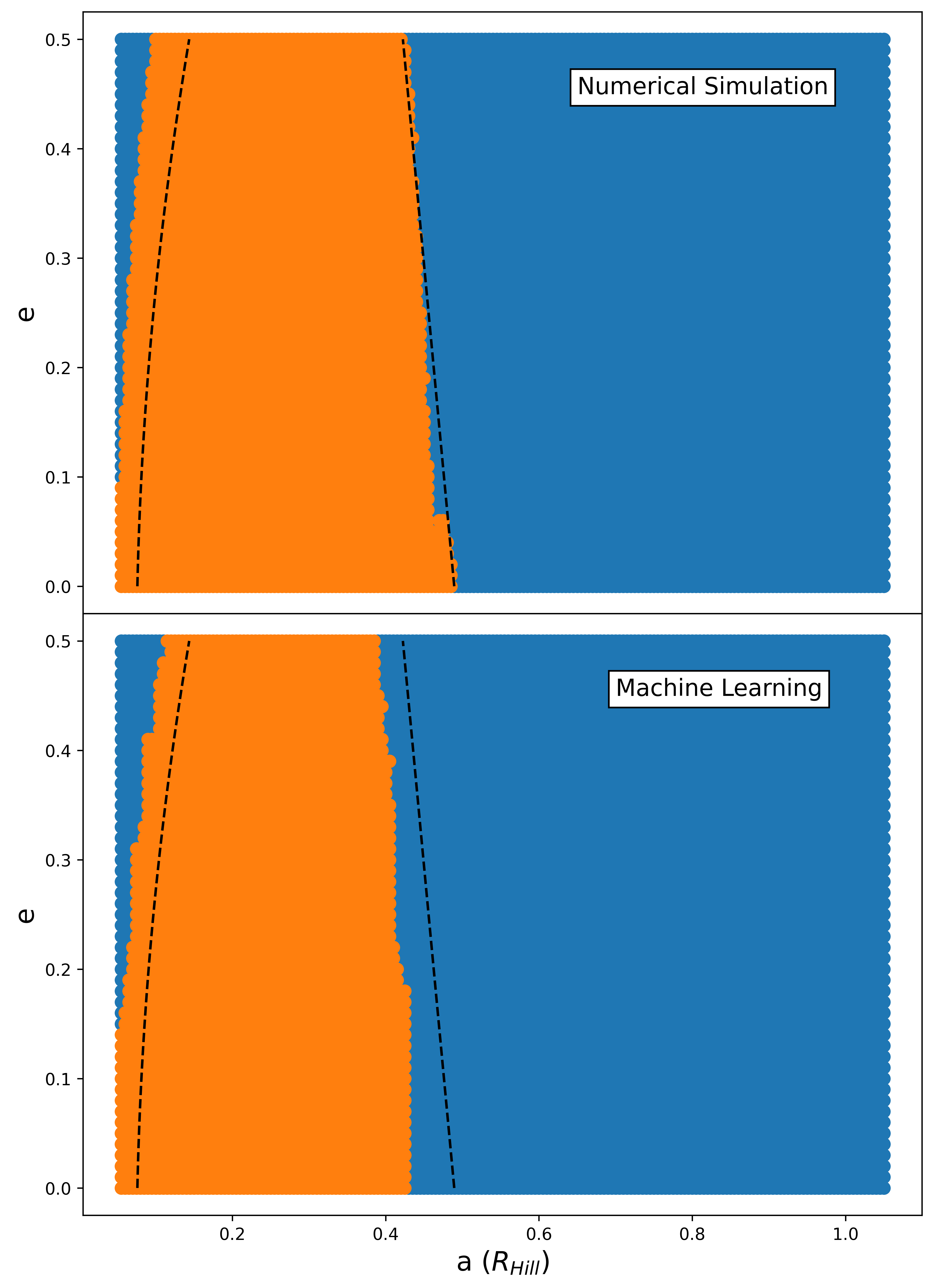}
  \end{subfigure}
  \hfill
  \begin{subfigure}{0.49\linewidth}
  \includegraphics[width=\linewidth]
  {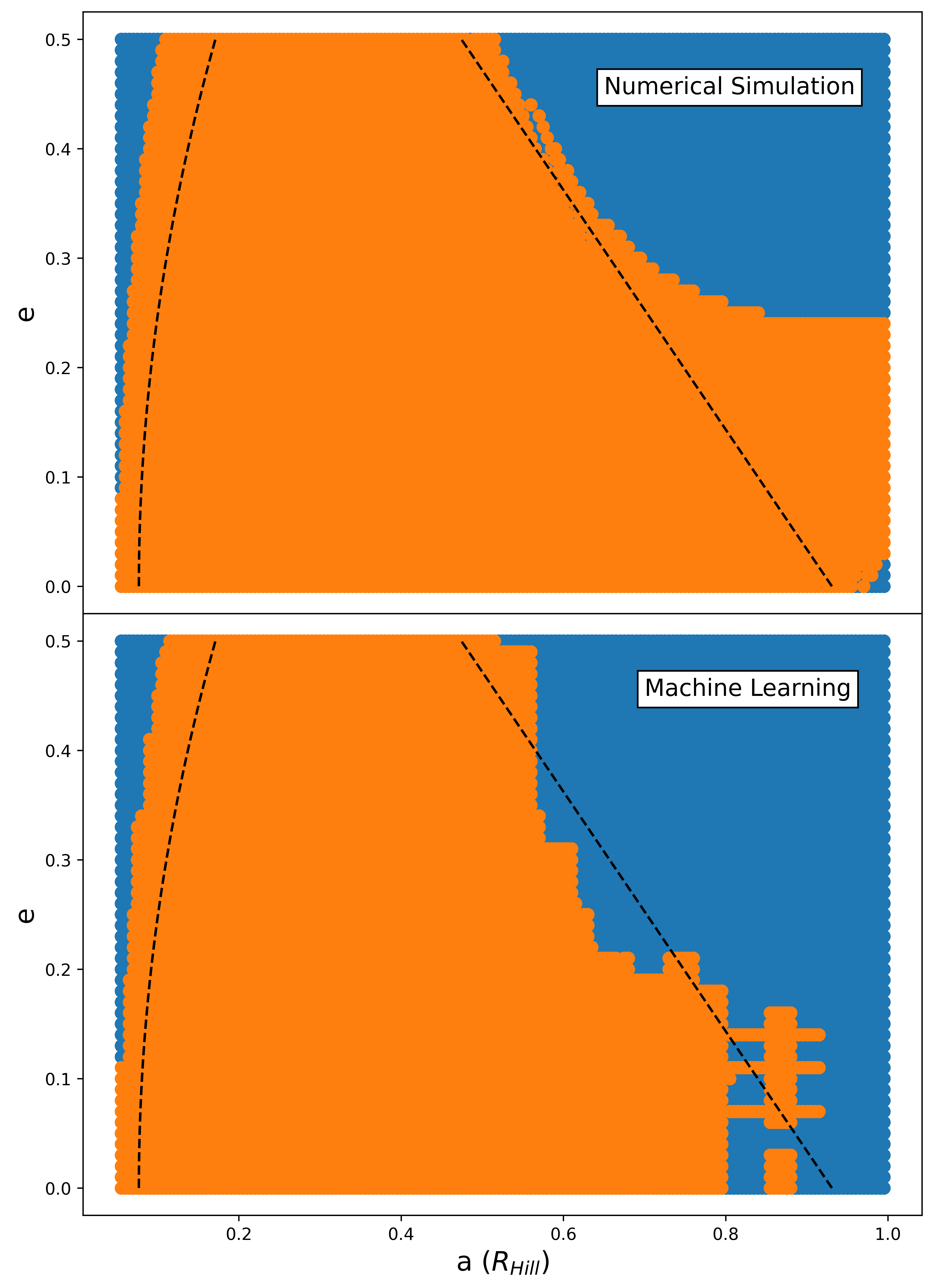}
\end{subfigure}
      \caption{Comparison of numerical simulations (upper panels) and ML predicted
outcomes (lower panels) for stable (orange) and unstable (blue) particles.
Initial conditions match those from \citet{Domingos2006} for $\mu = 10^{-3}$,
with i = 0° (left) and i = 180° (right).
The dashed lines represent the analytical expression from \citet{Domingos2006}.}
\label{rita}
\end{figure*}

\begin{figure}[!h]
    \centering
   \resizebox{0.47\textwidth}{!}{\includegraphics{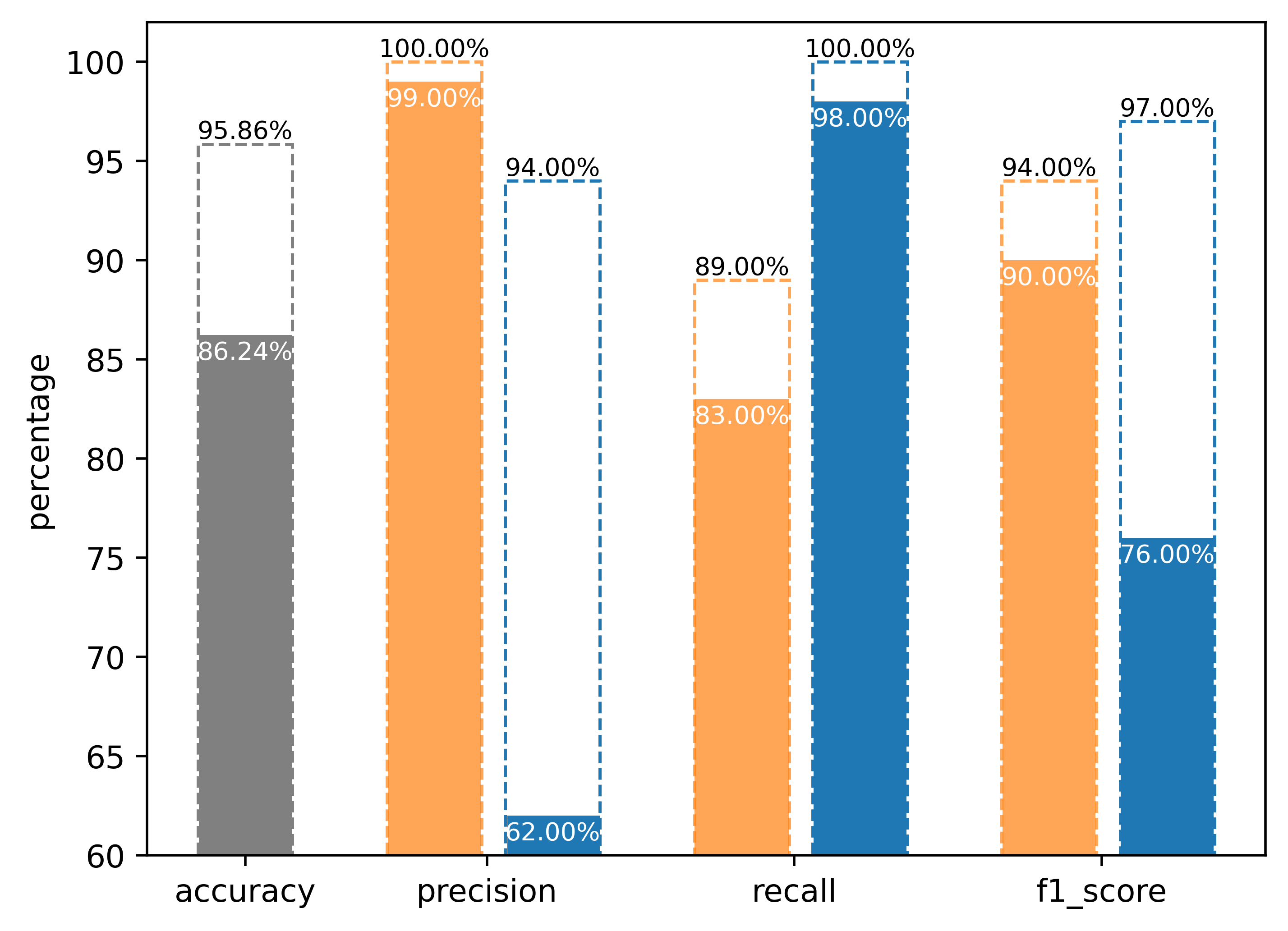}}
        \caption{Performance comparison between numerical simulations and machine
learning predictions for prograde (dashed line) and retrograde (solid line)
orbits from \citet{Domingos2006}. Blue and orange regions represent
unstable and stable particles, respectively.}
    \label{rita_performance}
    \end{figure}

Figure \ref{rita} compares the numerical simulation results with machine
learning predictions, with the left panels representing the results
for a prograde system and right panels for retrograde ones. The performance
of our model predictions are summarized in figure \ref{rita_performance},
where the dashed lines refers to a prograde system and the 
solid lines to the retrograde case.
The dashed lines represent the analytical expression
derived by \citet{Domingos2006}.

In both scenarios, the model accurately classified the main structure of the
stable region (orange points), but showed some discrepancies at the
boundaries. For the prograde case, the model closely replicated the stable
region, with only minor differences at the edges. The stable region extends to
about 0.49 $r_\text{Hill}$ in both the numerical and ML results, aligning
well with Domingos et al. (2006). The model achieved an overall accuracy of
95.86\%,  a precision of 94\% for unstable particles and 100\% for stable
particles, a recall of 100\% for unstable 
particles and 89\% for stable particles, 
and a F1-score of 97\% for unstable particles and 94\% for stable particles.

The performance comparison between the analytical expression 
for the critical semi-major axis proposed by \citet{Domingos2006} 
and our ML model reveals similar overall results. 
While \citet{Domingos2006}'s equations exhibit marginal higher accuracy (96.05\%) 
and recall for the stable and unstable classes (91.80\% and 98.70\%, respectively), 
our ML model outperforms in precision for both classes. 
Specifically, \citet{Domingos2006} achieves a precision of 97.78\% 
for stable particles and 95.08\% for unstable particles, 
whereas our model achieves 100\% precision in both classes.
The F1-score for both methods have the same percentage.

For the retrograde case, the model captured the overall larger stable
region, extending to about 0.93 $r_\text{Hill}$. However, it slightly
underestimated the stable region's extent for eccentricities above 0.3. The ML
model also missed some stable particles with semi-major axes beyond
0.8 $r_\text{Hill}$, particularly at lower eccentricities. 

Despite these limitations, the ML model achieved an accuracy of 86.24\%, 
outperforming the analytical expression by \citet{Domingos2006}, 
which had an accuracy of 85.04\%. 
Furthermore, the ML model attained a precision of 99\% for stable particles and 
62\% for unstable particles, a recall of 83\% for stable particles and 98\% for unstable particles. 
These results yielded F1-scores of 90\% for stable particles and 76\% for unstable particles. 
In comparison, \citet{Domingos2006}'s equations achieved slightly lower F1-scores, 
with 89.26\% for stable particles and 74.85\% for unstable particles, 
along with a recall of 80.81\% for the stable class and a precision of 59.94\% 
for the unstable class. 
However, their precision and recall for stable and unstable particles 
were marginally better, with values of 99.83\% and 99.54\%, respectively.

These results demonstrate the model's ability to accurately predict the
general structure of stable regions for both prograde and retrograde orbits,
while highlighting areas for potential improvement, particularly in
capturing fine details at stability boundaries and for highly eccentric
orbits in retrograde systems. 
Even the analytical expression by \citet{Domingos2006}, 
which applies only to $\mu$ = $10^{-3}$ and inclinations of 0$^\circ$ or 180$^\circ$
failed to classify many stable particles beyond the black line.
Further refinement of the model, possibly
through an expanded dataset or enhanced feature engineering, could address
these minor discrepancies.

\subsection{Comparison with \citet{Pinheiro2021}}

\citet{Pinheiro2021} studied the stability of a possible ring
system around the candidate planet PDS110b.
They ran $1.3 \times 10^6$ numerical simulations of
the three-body problem, each representing a different initial condition.
We evaluated the performance of our model for the PDS110b system,
specifically considering a scenario where the planet
has a mass equivalent to 6.25 Jupiter masses, an eccentricity of 0.2,
and a ring inclination of $150^\circ$.
This scenario corresponds to one of the possible parameters for
a retrograde ring system around PDS110b.

\begin{figure}[!h]
\centering
\resizebox{\hsize}{!}{\includegraphics{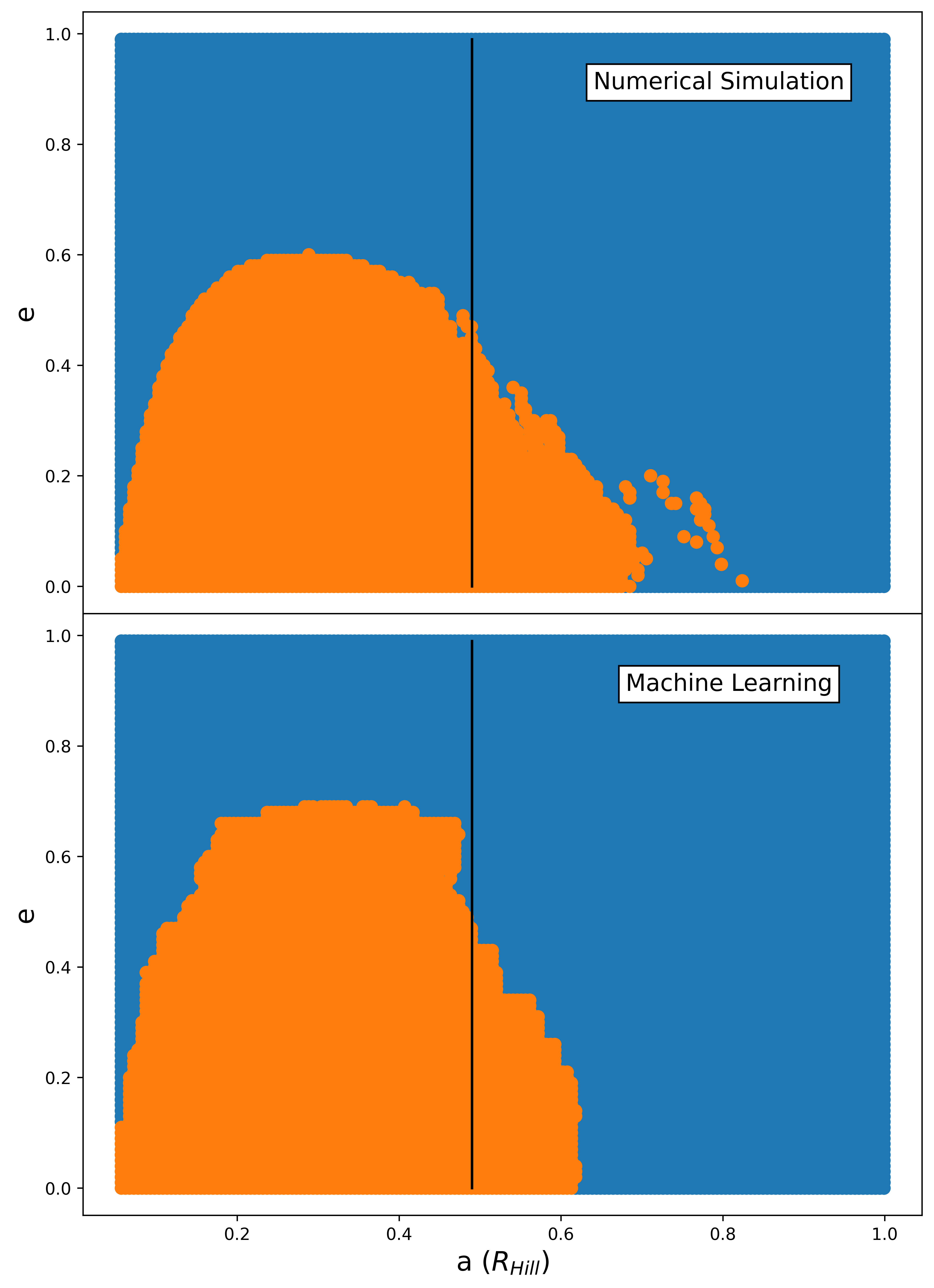}}
\caption{Stability maps for PDS110b. Top panel: Numerical simulation results.
Bottom panel: ML prediction. The black line represents the
minimum radial size for the observed ring as reported by \citet{Pinheiro2021}.}
\label{PDSSaturn}
\end{figure}

Figure \ref{PDSSaturn} shows the numerical and predicted stability maps around 
PDS110b. Although this system is retrograde, in this case
the model performed much better than in the previous example,
achieving an accuracy of 93.68\%
over $\sim$ 20,000 predicted initial conditions.
Here, the number of stable and unstable instances is almost
equivalent, with 5,722 actual stable and
14,373 actual unstable instances, respectively.

\begin{figure}[!h]
\resizebox{0.47\textwidth}{!}{\includegraphics{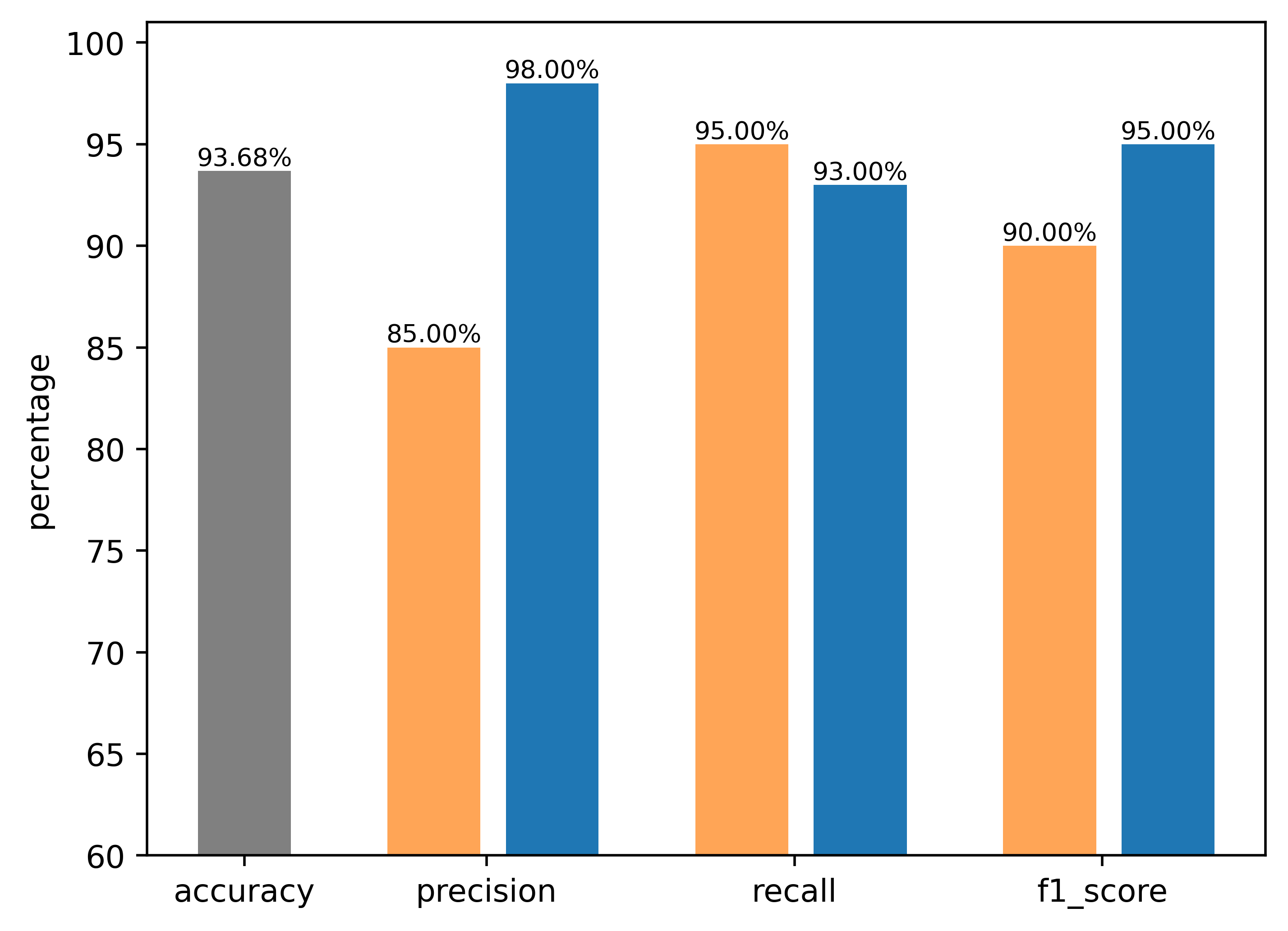}}
\caption{Performance metrics for the PDS110b system prediction.}
\label{PDS_performance}
\end{figure}

Figure \ref{PDS_performance} shows
precision, recall and F1 scores for both classes,
ranging between 85\% and 98\%.
The slight difference is attributed to the model's generalization
at the boundary of the stable region,
where it misclassified 286 instances as unstable
and 984 instances as stable.
Some of these misclassifications belong
to outlier particles outside the stable region.
Our numerical simulations cover a time span of $10^4$ orbital periods;
however, some of these stable particles might become unstable
with longer integration times.

\subsection{Stability map of Saturn’s Inuit satellites}
Some irregular Saturn moons are classified into three different groups
distinguished by their inclination range.
We generated a stability map for the Inuit group, where
the range of inclination is between $40^\circ$ and $50^\circ$.
Table \ref{moons} summarizes the orbital 
elements of the moons belonging to this group. 

\begin{table}[ht]
\caption{Orbital elements of Saturn satellites belonging to the Inuit group.
Some recently discovered Inuit satellites were not included
in this table because their orbital eccentricities are not well determined (https://carnegiescience.edu/).}
\label{moons}
\begin{center}
\begin{tabular}{llll}
\hline
Moon      & a ($r_\mathrm{Hill}$) & e   & i ($^\circ$) \\
\hline
Kiviuq    & 0.1803                                    & 0.334 & 45.71          \\
Ijiraq    & 0.1805                                    & 0.316 & 46.44          \\
S/2019 S1 & 0.1825                                    & 0.623 & 44.38          \\
Paaliaq   & 0.2466                                    & 0.364 & 45.13          \\
Siarnaq   & 0.2844                                    & 0.295 & 45.56          \\
Tarqeq    & 0.2922                                    & 0.160 & 46.09          \\
S2428b    & 0.2835                                    & 0.472 & 44.43          \\
T522499   & 0.2824                                    & 0.242 & 48.11         \\
\hline
\end{tabular}
\end{center}
\end{table}

As another test, we run $\sim$ 20,000 numerical simulations 
of three body problem representing Sun, Saturn 
(corresponding to {$\mu = 2.857\times 10^{-4}$}) and test particle.
The test particle's initial orbital inclination was set to $45^\circ$,
with the argument of pericenter, true anomaly, and longitude of node being randomly
selected from $0^\circ$ to $360^\circ$ using a uniform
distribution.

\begin{figure}[!h]
\centering
\resizebox{\hsize}{!}{\includegraphics{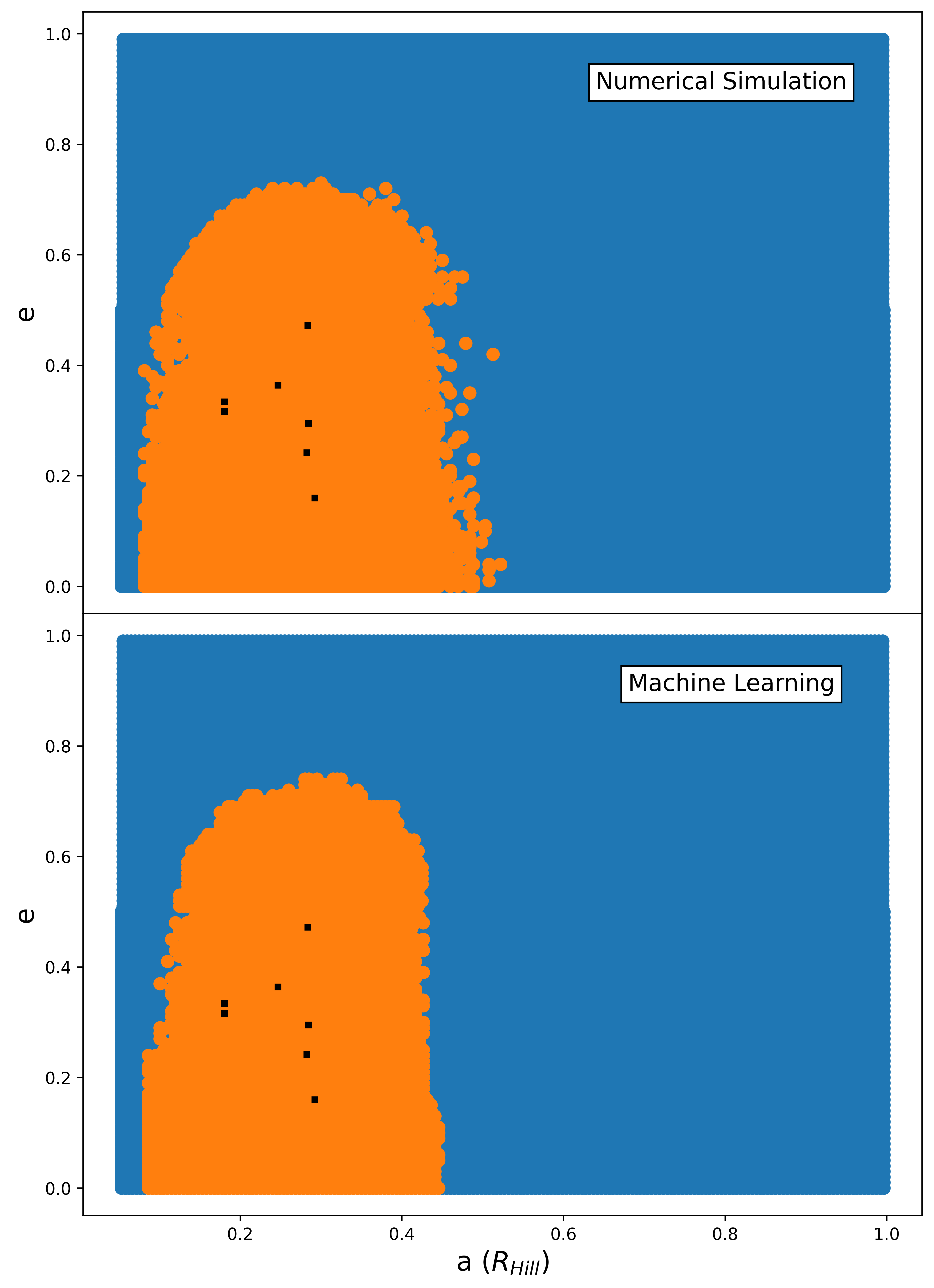}}
\caption{Stability maps for Saturn's Inuit satellites. Top panel: Numerical simulation results.
Bottom panel: ML prediction. 
Black squares represent the actual satellites
belonging to the Inuit group.}
\label{F-inuit}
\end{figure}

Figure \ref{F-inuit} shows the numerical and machine
learning predictions for this example.
The black squared points represent the satellites
belonging to the Inuit group.

\begin{figure}[!h]
\resizebox{0.47\textwidth}{!}{\includegraphics{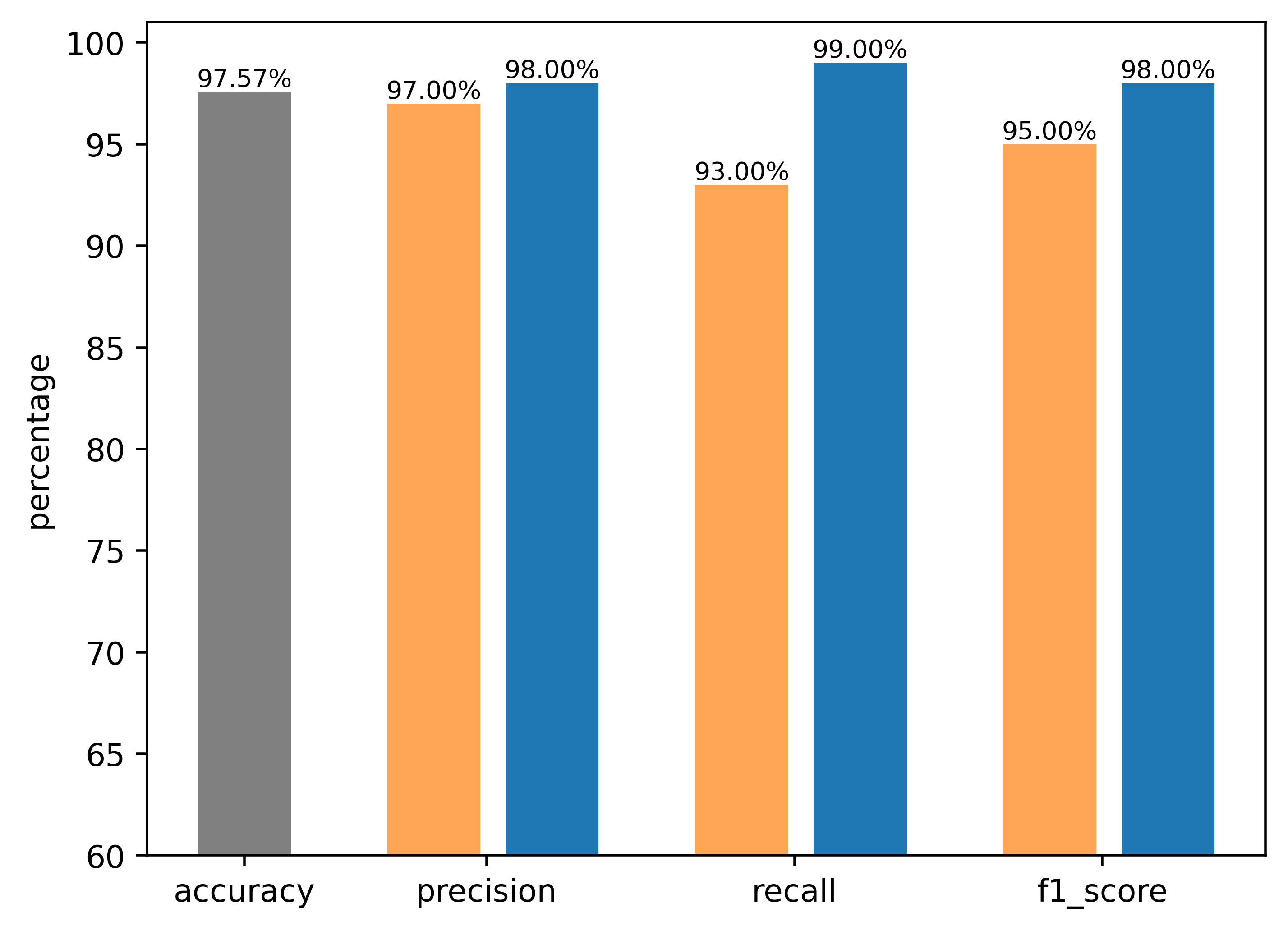}}
  \caption{Performance metrics for the Saturn's Inuit satellites prediction.}
\label{F-inuit_performance}
\end{figure}

This example produced our best performance with an
accuracy of 97.57\%, along with a precision and
recall for the stable class of 97\% and 93\%,
respectively, and with a precision and recall 
for the unstable class of 98\% and 99\% (Fig.~\ref{F-inuit_performance}).
In total, the ML model incorrectly
classified only 470 instances out of $\sim$ 20,000,
with most of them being spread particles at the
boundary of the stable region.
This good result is reflected in the F1-score,
which for both classes reached high values,
with 95\% for the stable class and 98\% for the unstable one.

\section{Final remarks}
\label{s-finalremarks}
This study demonstrates the efficacy of ML techniques in predicting
orbital stability for hypothetical planet.
We utilized a comprehensive dataset derived from $10^5$ numerical simulations of
three-body systems, encompassing a wide range of orbital and physical parameters.
These parameters included the mass ratio of the system, semi-major axis,
inclination, argument of pericenter, longitude of the node, eccentricity, and
true anomaly for both the planet and test particles.

Our numerical simulations revealed that 11.83\% of particles remained stable
throughout the integration period, while 47.22\% collided with the planet and
40.95\% were ejected from the system. This imbalanced class distribution
necessitated the application of resampling methods to enhance model performance.
We evaluated five ML algorithms: Random Forest, Decision Tree,
XGBoost, LightGBM, and Histogram Gradient Boosting,
all algorithms demonstrated comparable accuracy performance.
Through rigorous hyperparameter tuning and threshold optimization, we identified XGBoost
as the best-performing model, achieving an accuracy of 98.48\%.
To validate our model's generalization capabilities, we applied it to reproduce
the numerical results of \citet{Domingos2006}, \citet{Pinheiro2021}, and a
stability map for Saturn's Inuit group of satellites. The model demonstrated
robust performance, achieving an accuracy of 97.57\% across these diverse
scenarios.
Even though our dataset covers inclinations from 
$0^\circ$ to $180^\circ$, near i=$180^\circ$ the stable region 
exhibits significant sensitivity to minor inclination variations, 
resulting in decreased performance. 
Additionally, we also tested the performance of our model 
in the case of Earth ($\mu$ = 3$\times10^{-6}$) 
and obtained an accuracy of 95.17\%.

Our results show the potential of ML algorithms as powerful
tools for reducing computational time in three-body 
simulations in order $10^{5}$ 
times faster than traditional numerical simulations.
The ability to generate stability maps covering a wide range of orbital and 
physical parameters
within seconds represents a significant advancement in 
efficiency compared to
traditional numerical methods. 
The implementation of our model will be made accessible 
through a public web interface, facilitating its use by the broader scientific community. 

As future work we are refining the model's performance,
particularly in capturing fine details at stability 
boundaries and for highly
eccentric orbits, and also highly  
retrograde system.

Additionally, we plan to expand the dataset to include more diverse
planetary system configurations could enhance 
the model's generalization
capabilities and broaden its applicability to a wider range of 
astrophysical scenarios.


\begin{acknowledgements}
    We would like to thank the anonymous referee 
for the constructive comments that greatly improved the 
manuscript.

This research was financed in part by: 
Coordenação de Aperfeiçoamento de Pessoal de Nível Superior - 
Brasil (CAPES) - Finance Code 001;  
Fundação de Amparo a Pesquisa no Estado de São Paulo 
(FAPESP) - Proc. 2016/24561-0; German Research Foundation (DFG) - Project 446102036.

\end{acknowledgements}


\bibliographystyle{aa}
\bibliography{aanda}


\begin{appendix}
\section{Machine learning algorithms}
\label{Ss-Algorithms}

Here we present the main concepts and 
characteristics of the tree-based ML algorithms 
utilized in this paper.
\subsection*{Decision Tree}
\label{Ss-Decisiontree}

Decision Tree, first introduced by \cite{Quinlan1986}, classifies instances by
splitting them into leaf nodes \citep{Tom1997}. This method builds a tree
formed by a sequence of several nodes, each with a specific rule
($x_\mathrm{i} < t_\mathrm{i}$), where $t_\mathrm{i}$ is a threshold value of
a single feature ($x_\mathrm{i}$) in the $i^{th}$ node \citep{Shalev2014}. 
The tree starts with the root node, where the data are split into two
new subsets and nodes, which can be either a leaf node or a decision node.
The leaf or terminal node is where the instance is classified, while the
decision node splits the sample again into two new subsets and
nodes. This process repeats until the end of the branch. 

The choice of feature 
and best threshold value for each node depends on the impurity of the
subsequent two subsets, which can be calculated by estimating the class
distribution before and after the split \citep{Geron2022hands}.

\subsection*{Bagging classifier: Random Forest  }
\label{Ss-Bagging}
Generating a set of classifiers can improve accuracy and robustness. When
this approach uses the same training algorithm for every predictor to train
different random subsets, the method is called bagging \citep{Geron2022hands}.
The Bagging classifier creates random subsets of the dataset and trains a
model on each subset \citep{Breiman1996}. This method returns a
combination using a voting scheme of predictions from all predictors.

The bagging algorithm for trees works by iteratively selecting a subset of
instances from a training set through bootstrapping, then fitting a
tree to these selected instances. Predictions for unseen instances are
made by averaging the predictions from all individual trees
\citep{Cutler2011}.

Random Forest, introduced by \citet{Breiman2001}, is an example of a
bagging classifier and an ensemble of Decision Trees. Random Forest
reduces the risk of overfitting by decreasing the correlation between
trees \citep{Shalev2014}. It trains numerous decision trees using
different subsets of the training data. The final prediction is determined
by the most voted outputs among these individual decision trees.

\subsection*{Boosting classifiers} \label{Ss-Boosting}

Boosting aims to improve the accuracy of any given learning algorithm.
This method combines several weak learners into a strong learner and
trains predictors sequentially, with each one attempting to rectify the errors
of its predecessor \citep{Geron2022hands}. It uses the residuals of the
previous model to fit the next model. The following subsections present an
overview of XGBoost, LightGBM, and Histogram Gradient Boosting.

\subsection*{XGBoost} \label{Ss-XGboost}
XGBoost (Extreme Gradient Boosting) is a decision tree ensemble developed
by \cite{Chen2016} based on the idea of Gradient-Boosted Tree (GBT) by
\cite{Friedman2001}. It computes the residual of each tree prediction,
which are the differences between the actual values and the model
predictions. These residuals update the model in subsequent iterations,
reducing the overall error and improving predictions
\citep{Wade2020hands}. Unlike sequentially built Gradient-Boosted Trees,
XGBoost splits the data into subsets to build parallel trees during
each iteration.

XGBoost differs from Random Forest in its training process. It
incorporates new trees predicting the residuals of previous trees, and
combines these predictions, assigning varying weights to each tree, for 
the final prediction \citep{Chen2016}.

\subsubsection*{LightGBM} \label{Ss-LightGBM}
Light Gradient Boosting Machine (or LightGBM) is a variant of the
Gradient-Boosted Tree released in 2016 as part of Microsoft's Distributed
Machine Learning Toolkit (DMTK) project \citep{Ke2017}. The method employs
histograms to discretize continuous features by grouping them into
distinct bins. LightGBM introduces the leaf-wise growth strategy, which 
converges quicker and achieves lower residual.

LightGBM provides different characteristics compared to XGBoost, 
offering alternative benefits that may be more suitable depending on the specific use case.
These include improved memory usage,
reduced cost of calculating the gain for each split, and reduced
communication cost for parallel learning, resulting in lower training time
compared to XGBoost. The algorithm also utilizes the new Gradient-Based
One-Sided Sampling (GOSS) and Unique Feature Bundle (EFB) techniques
\citep{Ke2017}. GOSS creates the training sets for building the base
trees, while EFB groups sparse features into a single feature
\citep{Bent2021}.

\subsubsection*{Histogram Gradient Boosting \label{Ss-HGB}}
Histogram Gradient Boosting is a boosting model inspired by
LightGBM. For training sets larger than tens of thousands of instances, this
technique, with its histogram-based estimators, proves very efficient
and faster than previous boosting methods \citep{Wade2020hands}. Traditional
decision trees require long processing times and heavy computation
for huge data samples, as the algorithm depends on splitting all
continuous values and characteristics \citep{Ibrahim2023}.

Histogram Gradient Boosting streamlines this process by using histograms to bin the continuous
instances into a constant number of bins. This approach reduces the number of
splitting points to consider in the tree, allowing the algorithm to
leverage integer-based data structures \citep{Wade2020hands}. Consequently,
this improves and speeds up the tree implementation.

\section{{Imbalanced class}}\label{S-Imbalaced}

Random oversampling involves replicating
instances randomly within
the dataset to achieve a balance in class distribution.
This approach may lead to overfitting,
and an alternative solution to avoid this
is to use the SMOTE.

\citet{Chawla2002} proposed oversampling the
minority class by generating "synthetic"
instances. 
Firstly, this method calculates the
difference in feature vectors between
an instance and its nearest neighbors.
In the second step, it multiplies this
difference by a random number ranging from 0 to 1,
and a new instance is subsequently
created by adding this result to the
features of the original instance.

Another oversampling technique based on SMOTE
is Borderline-SMOTE \citep{Han2005}.
This method attempts to identify the borderline of
each class and avoids using the
outliers (noise points) of the minority
class to resample them. The algorithm works with the following steps:
\begin{enumerate}
\item For each instance of the minority class, check
which class its nearest neighbors belong to.
\item Count the number of nearest instances
that belong to the majority class.
\item If all neighbors are from other classes,
this point is classified as a noise
point, and it is ignored when resampling the data.
\item If more than half of the neighbors are from other classes,
this is a border point.
In this case, the algorithm identifies the $K$
nearest neighbors that are of the same class
and generates a synthetic instance between them.
\item If more than half of the neighbors are from the minority class,
it is a safe point and the SMOTE technique is applied normally.
\end{enumerate}

The last oversampling approach implemented
in this work is ADASYN,
which generates synthetic instances for
the minority class by considering
the weighted distribution of this particular class.

The methodology employed by \citet{He2008} involves an estimation of the
impurity of the nearest neighbors $r_{i}$ for each instance in the minority
class, given by 

\begin{align}
r_{i} = \frac{\Delta_{i}}{K}
\label{impurity}
\end{align}

\noindent where $\Delta_\mathrm{i}$ is the number of non-minority and $K$ the number of neighbors.

The subsequent step normalizes the $r_\mathrm{i}$ value as

\begin{align}
\hat{r_\mathrm{i}} = \frac{r_\mathrm{i}}{\sum\limits_{i=1}^{m} r_\mathrm{i}}
\label{normalized}
\end{align}

\noindent where $m$ is the size of the minority class data. This normalized value
is then multiplied by the total number of synthetic instances to be
generated. This procedure proportions the number of synthetic instances to
be created for each minority instance.
\end{appendix}

\end{document}